\documentclass{ar-1col-S2O}

\usepackage[comma]{natbib}
\usepackage{url}
\usepackage{xspace}
\usepackage{amsmath}
\usepackage{amssymb}
\usepackage{siunitx}
\usepackage{xspace}
\usepackage[T1]{fontenc}
\usepackage{csquotes}
\usepackage[hidelinks]{hyperref}

\newcommand{\arcsec}{\mbox{$^{\prime\prime}$}}% 

\newcommand{\eqns}[1]{\hyperref[#1]{Equations}~\ref{#1}}

\newcommand{\CD}{\ensuremath{C_\mathrm{D}}\xspace}
\newcommand{\alphamm}{\ensuremath{\alpha_\mathrm{mm}}\xspace}
\newcommand{\amax}{\ensuremath{a_\mathrm{max}}\xspace}
\newcommand{\cs}{\ensuremath{c_\mathrm{s}}\xspace}
\newcommand{\eps}{\ensuremath{\epsilon}\xspace}
\newcommand{\zed}{\ensuremath{Z}\xspace}
\newcommand{\hd}{\ensuremath{h_\mathrm{d}}\xspace}
\newcommand{\hg}{\ensuremath{h_\mathrm{g}}\xspace}
\newcommand{\Kn}{\ensuremath{\mathrm{Kn}}\xspace}
\newcommand{\kB}{\ensuremath{k_\mathrm{B}}\xspace}
\newcommand{\lambmdamfp}{\ensuremath{\lambda_\mathrm{mfp}}\xspace}
\newcommand{\Mstar}{\ensuremath{M_\star}\xspace}
\newcommand{\om}{\ensuremath{\Omega_\mathrm{K}}\xspace}
\newcommand{\Rep}{\ensuremath{\mathrm{Re}_\mathrm{p}}\xspace}
\newcommand{\Ret}{\ensuremath{\mathrm{Re}}\xspace}
\renewcommand{\mp}{\ensuremath{m_\mathrm{p}}\xspace}
\newcommand{\rc}{\ensuremath{r_\mathrm{c}}\xspace}
\newcommand{\reff}{\ensuremath{r_\mathrm{eff}}\xspace}
\newcommand{\rhod}{\ensuremath{\rho_\mathrm{d}}\xspace}
\newcommand{\rhog}{\ensuremath{\rho_\mathrm{g}}\xspace}
\newcommand{\rhos}{\ensuremath{\rho_\mathrm{s}}\xspace}
\newcommand{\Sigd}{\ensuremath{\Sigma_\mathrm{d}}\xspace}
\newcommand{\sigd}{\ensuremath{\sigma_\mathrm{d}}\xspace}
\newcommand{\Sigg}{\ensuremath{\Sigma_\mathrm{g}}\xspace}
\newcommand{\Sc}{\ensuremath{\mathrm{Sc}}\xspace}
\newcommand{\St}{\ensuremath{\mathrm{St}}\xspace}
\newcommand{\ts}{\ensuremath{t_\mathrm{stop}}\xspace}
\newcommand{\torb}{\ensuremath{t_\mathrm{orb}}\xspace}
\newcommand{\tgrow}{\ensuremath{t_\mathrm{grow}}\xspace}
\newcommand{\tdrift}{\ensuremath{t_\mathrm{drift}}\xspace}
\newcommand{\tdiff}{\ensuremath{t_\mathrm{diff}}\xspace}
\newcommand{\tsett}{\ensuremath{t_\mathrm{sett}}\xspace}
\newcommand{\vrel}{\ensuremath{v_\mathrm{rel}}\xspace}
\newcommand{\vsett}{\ensuremath{v_\mathrm{sett}}\xspace}
\newcommand{\vfrag}{\ensuremath{v_\mathrm{frag}}\xspace}
\newcommand{\vth}{\ensuremath{\bar v_\mathrm{th}}\xspace}
\newcommand{\vk}{\ensuremath{v_\mathrm{K}}\xspace}

\jname{Annu. Rev. Astron. Astrophys.}
\jvol{(62 / page)}
\jyear{2024}
\begin{document}

% Page header
\markboth{Birnstiel}{Growth and evolution of dust in protoplanetary disks}

% Title
\title{Dust growth and evolution in protoplanetary disks}

%Authors, affiliations address.
\author{Tilman Birnstiel
    \affil{University Observatory, Faculty of Physics, Ludwig-Maximilians-Universität München, Scheinerstr. 1, 81679 Munich, Germany;}
    \affil{Exzellenzcluster ORIGINS, Boltzmannstr. 2, D-85748 Garching, Germany;\\email: til.birnstiel@lmu.de}
}

%Abstract
\begin{abstract}
    %Abstract text, approximately 225 words and 
    %inclusive of 3--5 bullet items describing 
    %the findings of the current research:
    Over the past decade, advancement of observational capabilities, specifically the Atacama Large Millimeter/submillimeter Array (ALMA) and SPHERE instrument, alongside theoretical innovations like pebble accretion, have reshaped our understanding of planet formation and the physics of protoplanetary disks. Despite this progress, mysteries persist along the winded path of micrometer-sized dust, from the interstellar medium, through transport and growth in the protoplanetary disk, to becoming gravitationally bound bodies. This review outlines our current knowledge of dust evolution in circumstellar disks, yielding the following insights:\\[0.5em]
    \begin{minipage}{\hsize}
        \begin{itemize}
    \item Theoretical and laboratory studies have accurately predicted the growth of dust particles to sizes that are susceptible to accumulation through transport processes like radial drift and settling.
    \item Critical uncertainties in that process remain the level of turbulence, the threshold collision velocities at which dust growth stalls, and the evolution of dust porosity.
    \item Symmetric and asymmetric substructure are widespread. Dust traps appear to be solving several long-standing issues in planet formation models, and they are observationally consistent with being sites of active planetesimal formation.
    \item In some instances, planets have been identified as the causes behind substructures. This underlines the need to study earlier stages of disks to understand how planets can form so rapidly.
        \end{itemize}
    \end{minipage}\\[0.5em]
    In the future, better probes of the physical conditions in optically thick regions, including densities, turbulence strength, kinematics, and particle properties will be essential for unraveling the physical processes at play.
\end{abstract}

%Keywords, etc.
\maketitle

%Table of Contents
\tableofcontents

%%%%%%%%%%%%%%%%%%%%%%%%%%%%%%%%%%%%%%%%%%%%%%%%%%%%%%%%%%%%%%%%%%%%%%%%%%%%%%%%%%%%%%%%%%%%%%%%%%%%%%%%%%%%%%%%

% =========================================================
\section{INTRODUCTION}

% -----------------------------------------------------------
\subsection{Setting the stage: planet formation}

Today, planetary systems can be studied in great detail, from studies of our own planet, meteoritics, and other solar system explorations, to exoplanets, and other planetary systems in formation. It is quite surprising, that despite this wealth of data, the question of how planets form remains unanswered on a very fundamental level. It is generally accepted that an accretion disk forms due to angular momentum conservation as a by-product of star formation. At typical ISM conditions, around 1\% of the mass of that disk is in sub-micrometer dust particles \citep{Weingartner2001}. These particles need to be brought together to form bodies of kilometers to thousands of kilometers in size. However, as \citet{Youdin2005} write, the problems occur embarrassingly early: are the first gravitationally bound objects, the planetesimals, formed by gradual collisional growth or by gravitational collapse of over-densities, or by a combination of both effects? To understand the challenges involved and ultimately answer this question, the collisional evolution of the solid particles and their dynamics and transport need to be understood. Both of these aspects are intimately linked to each other: the particle size and composition determine the aerodynamic properties of the particle. In turn, the transport and dynamics determine collision speeds and how particles can be locally accumulated. Furthermore, the evolution of the particles depends on many other unknowns that can fundamentally change the picture: this ranges from the unknown source of turbulence (or lack thereof) to the unknown porosity evolution of particles to the unknown mechanism that turns porous dust aggregates into igneous pieces of rock called chondrules as they are found in meteorites \citep{Connolly2016}.

\begin{marginnote}[]
    \entry{Planetesimals}{building blocks of planets, defined as bodies bound together by their own gravity.}
    \entry{Chondrules}{millimeter-sized, spherical inclusions that make up most of the mass in a family of rocky asteroids called \textit{chondritic meteorites}.}
    \entry{MRI}{Magnetorotational Instability.}
    \entry{VSI}{Vertical Shear Instability.}
\end{marginnote}

The role dust particles play in the formation of planetesimals goes far beyond providing the material for planet formation. Dust also plays a major role in determining the structure, evolution, and chemical composition of its parent accretion disk. Firstly, dust particles are the primary source of continuum opacity. They determine where starlight is absorbed or scattered and how the disk is heated, effectively setting the hydrostatic structure of the disk wherever dust is abundant \citep[see, for example][]{Gorti2009,Woitke2009}. In the same way, it determines the UV flux inside the disk which modulates heating, freeze-out and photo-destruction processes of gas-phase chemical species \citep{Jonkheid2004,Aikawa2006}. Secondly, the dust particles provide the surface area on which complex surface chemistry takes place \citep[e.g.,][]{Garrod2006}. As will be described later in this chapter, settling and radial drift of dust particles can also act as a conveyor belt for the volatile species that are formed or frozen-out on the dust surfaces, causing a large-scale redistribution of abundances within the disk \citep[e.g.,][]{Stepinski1997,Cyr1998,Ciesla2006,Krijt2016c,Stammler2017}. Furthermore, electrons can efficiently be captured on the grain surfaces, which alters the ionization state of the disk \citep[e.g.,][]{Okuzumi2009a,Ivlev2016} and influences the gas dynamics by allowing or preventing the MRI to develop \citep{Terquem2008,Okuzumi2012b,Delage2022}. Several other hydrodynamic instabilities such as the VSI depend on the local cooling timescale, which is also determined to a large part by the dust properties \citep[e.g.,][]{Lin2015,Malygin2017,Pfeil2019}.

Finally, starlight scattered by small dust in the disk surface as well as the thermal continuum emission of the dust particles themselves are the most readily available observational tracers that allow us to observe the disks within which the planets form, as will be discussed in the following section. All the points above underline the fact that dust, while only contributing about 1\% to the total mass of the system, is a crucial ingredient affecting all aspects of planet formation.

% -----------------------------------------------------------
\subsection{The curtain opens: planet-forming disks observed}
\label{sec:curtain_opens}

\begin{figure}[h!]
    \centering
    \includegraphics[width=\textwidth]{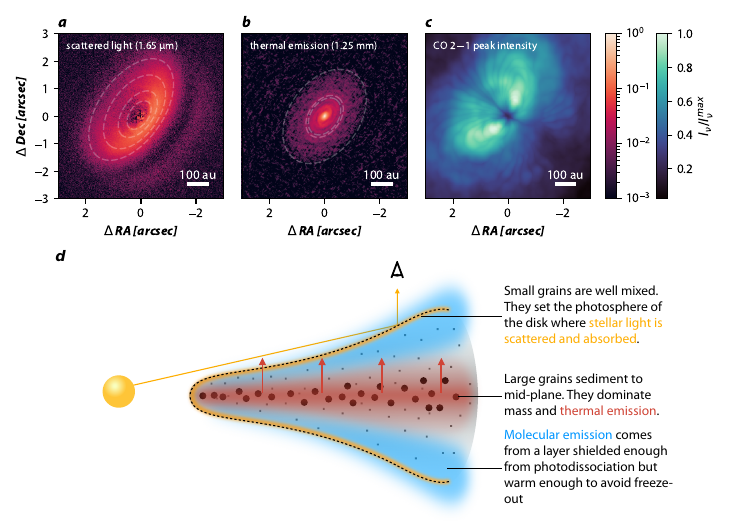}
    \caption{Observations for the disk around IM Lupi in scattered light \citep[][panel a]{Avenhaus2018}, thermal emission \citep[][panel b]{Andrews2018}, and carbon-monoxide 2-1 peak emission \citep[][panel c]{Oberg2021,Czekala2021}, and a schematic of the associated emission mechanisms (panel d). Circles in panel a are denoting surface features discussed in \citet{Avenhaus2018} while circles in panel b denote the rings fitted in \citet{Huang2018} and a \SI{1.6}{arcsec} circle denoting the outer edge of the continuum emission.}
    \label{fig:imlup}
\end{figure}

\begin{marginnote}
    \entry{SED}{Spectral Energy Distribution}
\end{marginnote}

Initial observations of planet-forming disks did not offer high enough resolution to spatially resolve the circumstellar material and were therefore limited to studying only the SEDs \citep[for a review, see, for example][]{Natta2007}. While extremely useful at the time, SED modeling was inherently limited due to the degenerate nature of the problem: Not every structural feature in the disk results in a unique spectral counterpart in the SED. Pioneered by the Hubble Space Telescope \citep[e.g.,][]{ODell1993,Debes2013}, high-resolution imaging of planet-forming disks has been revolutionizing the field. During the last decade, this was foremost driven by two observatories: firstly, the Atacama Large Millimeter/submillimeter Array (ALMA), a 66-antenna radio interferometer with baselines of up to 16 km, achieving unprecedented resolution and sensitivity for observing the long wavelength emission of disks. Around the time that the ALMA long baselines became available, also the VLT/SPHERE instrument \citep{Beuzit2008,Beuzit2019} saw first light, enabling high-contrast imaging of planet-forming disks in scattered light. Achieving comparable resolution (tens of milliarcseconds), ALMA and VLT/SPHERE can resolve \SI{1}{au}-scale structures in nearby disks \citep[e.g.,][]{Andrews2016,vanBoekel2017}. As an example, \autoref{fig:imlup} shows the same disk around the young star IM Lup in scattered light \citep[][H-band \SI{1.625}{\micro m}, left]{Avenhaus2018} as well as dust continuum \citep[\SI{1.25}{mm},][]{Andrews2018} and gas line emission (CO 2--1, \citealp{Oberg2021,Czekala2021}). In  panel d of \autoref{fig:imlup}, the emitting regions of each mechanism are shown schematically: the scattered light image is detecting stellar radiation that scatters off small dust particles in the disk surface due to the high optical depth of the disk at those wavelengths. The structure seen in the image depicts a vertically extended disk, as shown by the non-concentric ellipses \citep{Avenhaus2018}. In contrast, the millimeter emission shows a completely different morphology: the radial extend of most of the emission is much smaller, spiral features are seen in the inner parts of the disk and the fact that the circles that trace bright and dark rings (from \citealp{Huang2018} and a \SI{1.6}{arcsec} ring) are concentric, indicates little to no vertical extend of the emission. Comparing the scattered light image (\autoref{fig:imlup}, left panel) to the CO peak intensity image (right panel), it can be seen that the radial and vertical extend of the small dust particles and the gas are comparable.

The stark differences between these images point towards significant evolution of the solid component and can help guide the way towards a better understanding of the early phases of planet formation. The following section introduces the dynamics of solid particles in disks which can help explain the features seen in the observations above. We will see that this will require the particles to have reached sizes at least a hundred times larger than their initial ISM size scale. \autoref{sec:coll_evol} will therefore explain the physics of dust particle growth and the combined effects of growth and transport will be discussed in \autoref{sec:global_pic}. In the final \autoref{sec:path_ahead}, the subsequent growth towards planetesimals, observational methods to trace these theories, and future directions will be discussed.

\begin{textbox}[h]\section{The standard disk}
    \label{box:standard_disk}
    Within this review, unless otherwise noted, we will consider a vertically isothermal disk with a radius-dependent temperature $T(r) \simeq \SI{200}{K} \, (r/\si{au})^{-1/2}$. The disk column density is $\Sigg(r) = \frac{M_\mathrm{disk}}{2\pi\,\rc^2}\,\frac{\rc}{r}\,\exp\left(-r/\rc\right)$ which results in a vertical gas density profile of $\rhog(r,z) = \frac{\Sigg(r)}{\sqrt{2\pi}\,\hg} \exp\left(-\frac{z^2}{2\,\hg^2}\right)$, where $z$ is the height above the mid-plane, $\hg = \cs/\om$ is the gas scale height, \cs the isothermal sound speed and $\om=\sqrt{\frac{\mathrm{G}\, \Mstar}{r^3}}$ the Keplerian frequency. We will assume a stellar mass of $\Mstar = \SI{0.5}{M_\odot}$, a disk mass of $M_\mathrm{disk} = \SI{0.05}{\Mstar}$, a characteristic radius of $\rc = \SI{30}{au}$, and mean molecular weight of 2.3 proton masses.
\end{textbox}

% =========================================================
\section{DUST DYNAMICS}
\label{sec:dustdynamics}

The dynamical evolution of dust grains around stars can be affected by many mechanisms including resonances, radiation pressure, solar wind interaction, magnetic fields, and others. Inside planet-forming disks, however, small particles are most strongly affected by their coupling to the gas via drag forces (but see \citealp{Owen2019}). In the following, the key concepts will be introduced, and it will be shown how the drag force leads to systematic motion of dust and how the turbulent gas can act as diffusion. For other aspects, such as dust dynamics in disk winds, binaries, or warped disks, the authors may refer to more in-depth literature \citep[e.g.][and others]{Sellek2021,Aly2021,Zagaria2023}.

% -----------------------------------------------------------
\subsection{Drag Forces}
\label{sec:dragforces}

\begin{figure}[h!]
    \centering
    \includegraphics[width=\textwidth]{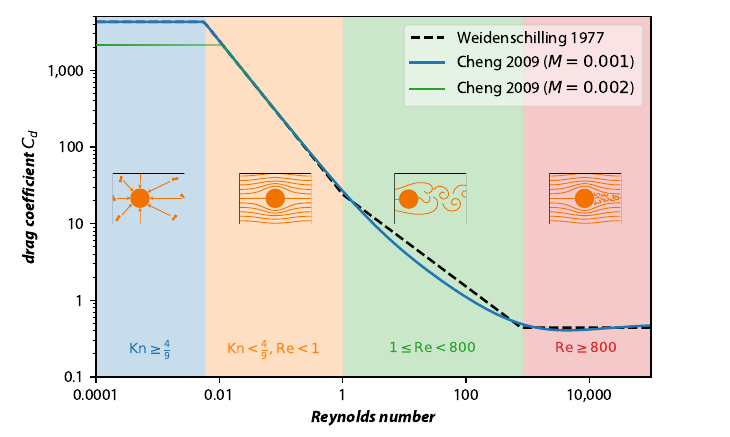}
    \caption{Drag coefficient and different drag regimes for subsonic conditions, adapted from \citet{Bagheri2016}. For Knudsen numbers larger than 4/9, the Epstein regime, the drag coefficient is inversely proportional to the dust-gas relative velocity which cancels out the velocity dependence in \autoref{eq:dragforce}. Here $M = \vrel/c_s$ denotes the Mach number.}
    \label{fig:drag}
\end{figure}

Particles moving with respect to the surrounding gas feel a drag force that acts to decelerate the particle until the relative velocity vanishes. Depending on the conditions the drag forces can be linear or non-linearly proportional to the magnitude of the relative velocity. For most regions in the disk and particles sizes accessible to observations, the mean free path of the gas molecules \lambmdamfp is larger than the particle radius $a$. This ratio $\Kn = \frac{\lambmdamfp}{a}$ is called the Knudsen number. On these scales, the gas molecules are not in the hydrodynamic limit. This means, the dust particle experiences the gas not as a flow, but as a bombardment of individual molecules, where the impact velocities of the molecules in the direction of motion of the particles are higher by the relative velocity between the dust and gas. This regime of the drag force is also called the Epstein regime after \citet{Epstein1924} (see also the box ``Drag Force Regimes''). If the particles become larger, their Reynolds number \Rep (see box "Turbulent Relative Velocities") increases while the Knudsen number decreases and the interaction between the dust particle and the gas becomes of a hydrodynamic nature, where the Reynolds number determines the flow structure, as shown in \autoref{fig:drag}.

By relating the momentum of the particle (relative to the gas) to the rate of momentum change (= the drag force), we can derive the timescale on which the particle velocity changes,
\begin{equation}
    \ts = \left| \frac{m\,\vrel}{F_\mathrm{D}}\right|.
\end{equation}
In the Epstein drag regime, the linear velocity dependence cancels out, and, if the particle is a sphere of density \rhos, the stopping time simplifies to
\begin{equation}
    \ts =  \frac{a \, \rhos  }{\vth  \, \rhog}.
\end{equation}
For conditions at \SI{1}{au} (see text box ``The standard disk''), the stopping time is only around a second for a micrometer-sized particle, but of the order of a month for a meter sized boulder. As conditions vary greatly throughout the disk, a more useful, dimensionless number is the Stokes number, $\St = \ts\,\om$, which relates the stopping time and the orbital timescale. Again assuming the Epstein drag regime, in the mid-plane, the Stokes number becomes
\begin{equation}
    \St = \frac{a\,\rhos}{\Sigg} \,\frac{\pi}{2},
    \label{eq:stokesnumber}
\end{equation}
so two particles with the same Stokes number behave aerodynamically identical, even if their shapes or compositions are different. \autoref{eq:stokesnumber} shows, that in the Epstein drag regime, the Stokes number is independent of velocity, and linearly dependent on the particle radius. This changes in the Stokes drag regime. Most relevant here is the Stokes drag at low Reynolds numbers, where the Stokes number is still velocity-independent, but scales as $a^2$. For larger Reynolds numbers, the stopping time is also velocity dependent.

\begin{textbox}[h]\section{Drag Force Regimes}
    The drag force for a dust particle of size $a$ moving at sub-sonic speed \vrel relative to the gas can generally be written as 
    \begin{equation}
        \label{eq:dragforce}
        \vec F_\mathrm{D} = - \frac{\CD}{2} \, \pi \, a^2 \, \rhog \, \vrel \, \vec\vrel,
    \end{equation}
    where \rhog is the gas density. \CD is the dimensionless drag coefficient which in the Epstein drag regime ($\Kn \geq 4/9$) is
    $$\CD^\mathrm{Epstein} = \frac{8 \, \vth}{3\, \vrel}.$$
    This results in a drag force that is linearly dependent on the relative speed with respect to the gas. Beyond the Epstein drag, the Stokes drag coefficient becomes dependent on the particle Reynolds number, $\Rep = 2\,a\, \vrel / \nu$, where $\nu$ is the gas viscosity.
    %turbulent reynolds number: $\Re = \frac{\alpha \Sigg \, \sigma_{H_2}}{2 \mu \, \mp}$
    The classical parameterization of \citet{Weidenschilling1977a} for the drag coefficient is
    \begin{equation}
        \CD \simeq
        \begin{cases}
            24   \, \Rep^{-1}   & \text{for } \Rep < 1                     \\
            24   \, \Rep^{-0.6} & \text{for } 1 < \Rep < 800               \\
            0.44               & \text{for } \Rep > 800                 
        \end{cases}
    \end{equation}
    The first two, \Rep-dependent cases are often called Stokes drag, while the last, constant value of \CD at large \Rep is sometimes called Newton drag. More recent empirical fits that better reproduce experimental results are, for example, given in \citet{Cheng2009} as
    \begin{equation}
        \CD = \frac{24}{\Rep}\, \left( 1 + 0.27 \, \Rep \right)^{0.43} + 0.47 \, \left[ 1 - \exp\left(-0.04 \, \Rep^{0.38}\right)\right]
    \end{equation}
    and compared in \autoref{fig:drag}.
    
\end{textbox}

% -----------------------------------------------------------
\subsection{Systematic Drift}

The dynamics of dust and gas are coupled via the drag forces, as described in the previous subsection. The amount of momentum lost by a decelerated dust grain is gained by the gas. Under nominal conditions, only about one percent of the mass is in solids \citep{Weingartner2001,Lodders2003}, so the dust experiences much stronger deceleration or acceleration than the gas. If the dust is locally enhanced and approaches a dust-to-gas mass ratio of unity, the dust and gas dynamics affect each other equally, which leads to complicated dynamics, as discussed later in this chapter.

A dust particle orbiting a star in vacuum will follow a Keplerian orbit. In a rest-frame that co-rotates with Keplerian velocity, this particle will oscillate vertically around the mid-plane (if it has non-zero inclination) and it will oscillate radially (if it has non-zero eccentricity), once per orbit. If the particle is now embedded in a gas disk that rotates \textit{exactly} Keplerian (which is not generally the case as we will see below), the radial and vertical oscillation of the particle will be damped by the drag forces. The orbital oscillations will change from a harmonic oscillation to a damped oscillation. The particle will approach an equilibrium once it has lost its eccentricity and inclination, so it will effectively have sedimented to a circular orbit at the mid-plane. At this point, it has reached the gas velocity and no drag forces remain.

For most particle sizes relevant to this review, the stopping time will be shorter than the orbital timescale ($\St < 1$) which means that particles approach the gas velocity before they have completed a single orbit. For the vertical sedimentation, this corresponds to an over-damped oscillation, similar to a falling feather that quickly approaches its \textit{terminal velocity}, where the velocity-dependent drag force becomes equal in magnitude to the gravitational acceleration. In the case of a sedimenting dust particle, the force balance requires $-z \,\om^2 - \frac{v_z}{\ts} = 0$, where the first term is the vertical component of the stellar gravitational acceleration and the second term is the deceleration due to vertical drag forces. This shows, that the vertical settling speed at height $z$ above the mid-plane is
\begin{equation}
    \vsett = - z \, \om \, \St,
    \label{eq:vsett}
\end{equation}
from which the timescale for sedimentation becomes $t_\mathrm{sett} = (\St\,\om)^{-1}$.

\begin{marginnote}
    \entry{Terminal velocity}{The velocity reached when all acceleration terms cancel out.}
\end{marginnote}

A more general derivation of the dust particle velocities under the influence of gas drag start from the Euler equations for a gas and a dust fluid of fixed Stokes number \citep[e.g.,][]{Youdin2005}. This approach allows analyzing the system for instabilities, as discussed in \autoref{sec:towards_planetesimals}. If we further ignore advective contributions, as in \citet{Nakagawa1986}, we arrive at 
\begin{align}
    \label{eq:dust_motion}
    \frac{\partial \vec v_d}{\partial t} & = - \frac{1}{\ts} ( \vec v_d - \vec v_g ) - \frac{\mathrm{G}\,M_\star}{r^3} \, \vec r                                \\
    \label{eq:gas_motion}
    \frac{\partial \vec v_g}{\partial t} & = - \frac{\eps}{\ts} (\vec v_g - \vec v_d ) - \frac{\mathrm {G}\,M_\star}{r^3} \, \vec r - \frac{\nabla P}{\rhog},
\end{align}
where $\eps = \rhod/\rhog$ is the dust-to-gas ratio.
It can be seen that a steady state is reached if
\begin{equation}
    \label{eq:terminalvelocity}
    \vec v_d - \vec v_g = \frac{\ts}{\rhog (1 + \eps)} \,\nabla P.
\end{equation}
This approximation, called the \textit{terminal velocity approximation} \citep{Youdin2005} shows that dust particles generally drift towards higher pressure. However, there are exceptions to this rule, for example if both the gas and the dust are on eccentric orbits. This breaks geostrophic balance and \autoref{eq:terminalvelocity} is not valid anymore. The lower velocity at apastron leads to a higher density. Yet, as both gas and dust have the same speed along their orbit, no drag forces are acting that accelerate or decelerate the dust particles along the orbit \citep[see,][]{Hsieh2012}.

Unlike to what was assumed above, planet-forming disks are pressure supported with a generally negative pressure gradient. In a dust-free environment, the force balance between gravity, centrifugal force, and pressure force leads to a slightly sub-Keplerian gas azimuthal velocity of 
\begin{equation}
    v_{g,\varphi} = \vk\,\sqrt{1 - 2 \, \eta} \simeq (1 - \eta)\, \vk,
\end{equation}
where
\begin{equation}
    \eta = -\frac{1}{2} \left(\frac{\hg}{r}\right)^2\,\frac{\partial \ln P}{\partial \ln r}.
    \label{eq:eta}
\end{equation}
For typical conditions, $\eta \sim \text{few} \times 10^{-3}\, \sqrt{r/\si{au}}$ (steeper in the exponential outer part of the disk), which means that the gas disk at \SI{1}{au} rotates very close to Keplerian speed, but the difference of several tens of \SI{}{m.s^{-1}} still causes significant drag on the dust particles.

The components of the gas and dust terminal velocity were derived in \citet[][for earlier works, see \citealp{Whipple1972} and \citealp{Weidenschilling1977a}]{Nakagawa1986} starting from \eqns{eq:dust_motion} and \ref{eq:gas_motion}. The resulting deviations from Keplerian velocity in polar coordinates are ($\vec u = (v_r, v_\varphi - \vk, 0)$)
\begin{align}
    \label{eq:nsh1}
    v_{d,r}       & = - \frac{2}{\St + \St^{-1} (1+\eps)^2}\,\eta\,\vk                                                \\
    v_{d,\varphi} & = - \frac{1 + \eps}{(1+\eps)^2 + \St^2}\,\eta\,\vk                                                \\
    v_{g,r}       & = - \frac{2 \, \eps}{\St + \St^{-1}\, (1+\eps)^2}\,\eta\,\vk                                      \\
    \label{eq:nsh4}
    v_{g,\varphi} & = \left(\frac{\eps}{1 + \eps} \cdot \frac{1}{1 + \St^2\,(1+\eps)^{-2}} - 1 \right)\,\eta \, \vk.
\end{align}
It can be seen that all velocities scale in magnitude with $\eta\,\vk$, the amount that the gas rotates faster or slower than the Keplerian speed. For low dust-to-gas ratio and $\St\ll1$, the dust radial speed becomes
\begin{equation}
    \label{eq:simpledrift}
    v_{d,r} = \St \left(\frac{\hg}{r}\right)^2 \frac{\partial \ln P}{\partial \ln r} \vk.
\end{equation}
For the values assumed here (see box \enquote{The Standard Disk}), this speed is about $\SI{-50}{m.s^{-1}} \simeq \SI{-10}{au.kyr^{-1}}$ and higher in the exponential part of the disk. This means that even particles of Stokes numbers of $10^{-2}$ can move through the entire disk within one million years.

\begin{figure}[h!]
    \centering
    \includegraphics[width=\textwidth]{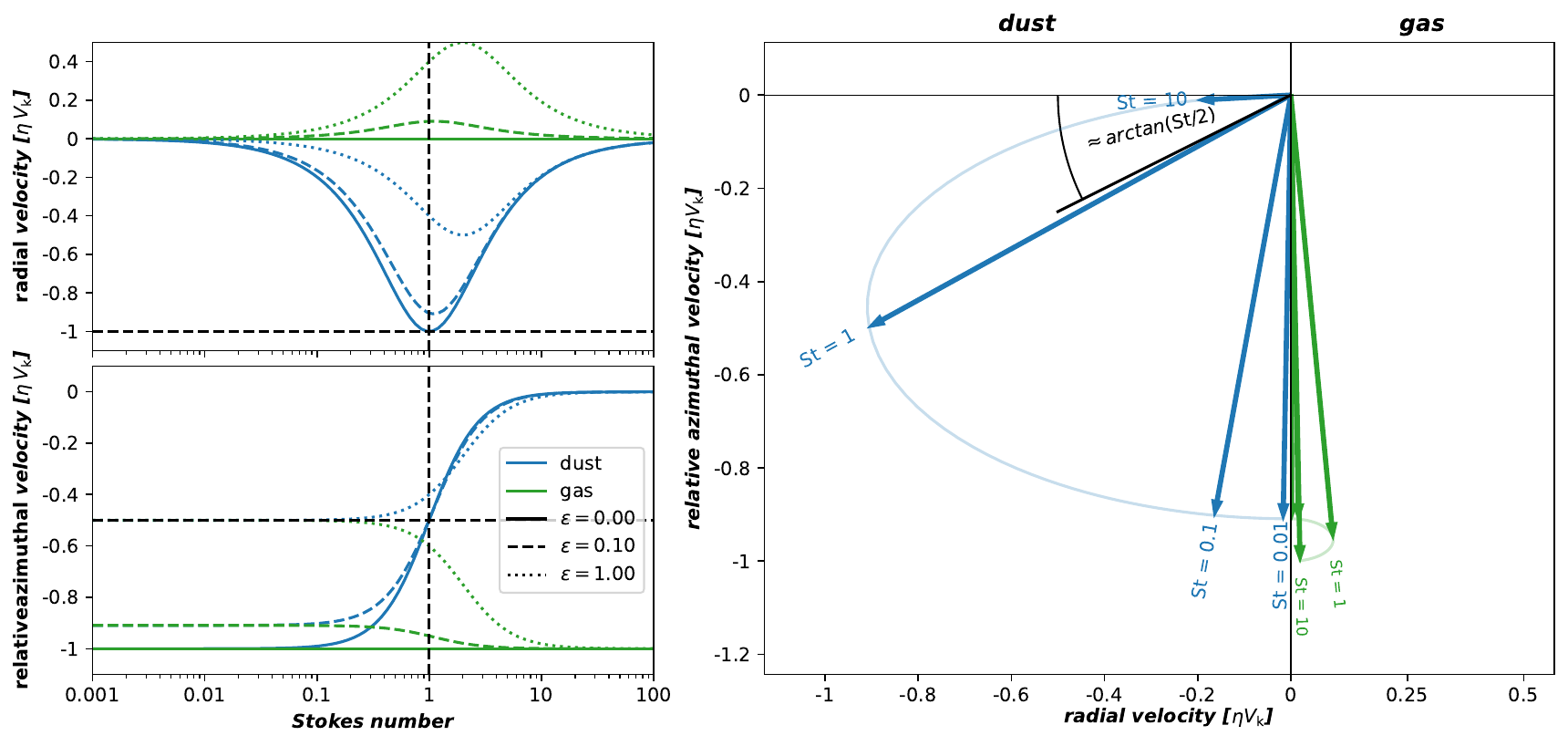}
    \caption{Dust and gas velocities in equilibrium \citep{Nakagawa1986}, displayed as deviations from Keplerian rotation normalized by $\eta \, \vk$. The left panels show the radial (top) and azimuthal (bottom) velocity components for the gas in green and dust in blue where solid lines denote a case without dust affecting the gas velocity, while dashed and dotted denote a dust-to-gas ratio of 0.1 and 1, respectively. The right panel shows the velocities as vectors for a dust-to-gas ratio of 10\%. It shows that dust feedback is strongest for $\St=1$, leading to a radial outfow of the gas which remains mostly $\eta\,\vk$ slower than Keplerian.}
    \label{fig:NSH}
\end{figure}

\eqns{eq:nsh1} through \ref{eq:nsh4}  also depend on the dust-to-gas ratio $\eps = \rhod/\rhog$. For the canonical value of 1\%, this effect is negligible and results in a substantial drift speed of the dust and no substantial drift of the gas. As \eps approaches unity, the radial drift of the dust will slow down and instead the gas will start to radially drift outward. \autoref{fig:NSH} shows the dust and gas velocity vectors and components, relative to Keplerian rotation for a dust-to-gas ratio of 10\% (to amplify the effects) and for different Stokes numbers. The maximum dust radial drift speed is reached when $\St=1$.  The radial drift of the gas is opposite in direction and lower by a factor of the dust-to-gas ratio, $v_{g,r} = - \eps v_{d,r}$. For small \eps and \St, both the gas and dust remains $\eta \vk$ slower than Keplerian. The necessary drag therefore comes from the fact that the dust has a radial component. In other words, the dust feels a head-wind that continuously extracts angular momentum. This headwind is not along the azimuthal direction, but at an angle of $\arctan(\St/2)$ (for $\eps\ll1$), so mostly radial for small \St and about 26$^\circ$ for $\St=1$. The dust approaches Keplerian speed only above $\St > 1$ and the gas only approaches Keplerian speed for dust-to-gas ratios above unity.

As discussed later, this classical picture of dust drift is complicated by several factors. Firstly, not all particles have the same size or Stokes number. A size distribution therefore needs to be accounted for in above equations which results in integrals in above equations of motion, complicating the resulting terminal velocities. See \citet{Tanaka2005}, \citet{Okuzumi2012a}, \citet{Dipierro2018}, or \citet{Garate2019,Garate2020} for details. Secondly, the dust-to-gas ratio can vary vertically if particles settle to the mid-plane. This will impose also a vertical stratification in the dust and gas radial speeds and the velocities used in vertically integrated simulations should account for that \citep[e.g.,][]{Garate2020}. As an example, gas pressure bumps with a high dust content would be disrupted if the velocities are computed from the vertically averaged quantities \citep{Taki2016}. If the vertical stratification is considered, then the impact of the sedimented dust is even stronger, but it is spatially limited to the mid-plane \citep{Onishi2017} and therefore cannot disrupt the entire vertical gas column. Similar conclusions were found for the disruption of vortices by \citet{Lyra2018}: two-dimensional considerations would predict disruption of vortices by dust feedback effects, but in 3D, these effects are limited to the mid-plane and do not strongly affect the vortex stability.

% -----------------------------------------------------------
\subsection{Turbulent Mixing}

The preceding section showed the systematic drift motion that particles experience in a near-Keplerian disk. However, on top of the ordered near-Keplerian shear flow, the gas might be turbulent. The coupling of the particles to the turbulent gas will also cause a random velocity component of the particles which leads to mixing/diffusion of the particle density and relative velocities between the particles. The latter is discussed in more detail in \autoref{sec:rel_vel}. In the following, we will discuss the effects of turbulent mixing.

Turbulent mixing of the dust is caused by the turbulent motion of the gas that is transferring its random motions to a certain amount to the particles \citep{Volk1980}. As an example, small particles that are tightly coupled to the gas quickly adapt to the motion of the gas. As they get picked up by turbulent eddies, they follow the motion of the eddy until it decays. Larger particles that might not be as strongly coupled to the gas flow still feel an acceleration due to their velocity difference with respect to the eddy velocity and a resulting drag force. In this latter case, a turbulent eddy induces a random kick to the particle momentum. The velocity dispersion of the dust can therefore be thought of a superposition of these effects integrated over all scales of the turbulent spectrum.

In planet-forming disks, the turbulent diffusivity is usually parameterized in terms of the Schmidt number \Sc. However, definitions vary between describing the ratio of diffusivity to viscosity of the gas $\nu/D_\mathrm{g}$, ratio of gas and dust diffusivity $D_\mathrm{d} / D_\mathrm{g}$, or combinations of those, see the discussion in \citet{Youdin2007}. In the following, we will assume that the gas viscosity and diffusivity are identical (i.e. mass is diffused in the same way as momentum), and will follow the definition of \citet{Youdin2007}, where the Schmidt number is
\begin{equation}
    \Sc = D_\mathrm{g} / D_\mathrm{d}.
\end{equation}

Within this picture of a turbulent cascade, \citet{Dubrulle1995} derived a Schmidt number of $\Sc = 1 + \St$ for vertical mixing. \citet{Youdin2007} generalized these considerations taking the effects of orbital motion into account: particles in orbit around a star diffuse differently than free bodies. The result of angular momentum conservation leads to a reduced  diffusivity of the particles of $\Sc = 1 + \St^2$. An extended derivation and discussion of this can be found in \citet{Binkert2023c}.

% -----------------------------------------------------------
\subsection{Drift / Mixing Equilibrium}
\label{sec:drift_mixing_equi}

The vertical evolution of a contaminant \citep{Morfill1984} under the influence of vertical settling and diffusion is described by an advection-diffusion equation,
\begin{equation}
    \frac{\partial \rhod(z)}{\partial t} + \frac{\partial}{\partial z} \left( \rhod\, \vsett - D \rhog \frac{\partial}{\partial z}\left( \frac{\rhod}{\rhog} \right) \right) = 0.
\end{equation}
A solution to this equation in steady state (and zero net vertical dust transport) is found where the diffusive (upward) flux is exactly opposite equal to the sedimentation (downward) flux. This solution is \citep[see][]{Fromang2009}
\begin{equation}
    \eps(z) = \eps(z=0) \, \exp\left( \int_0^z{\frac{\vsett(z')}{D(z')}} dz' \right),
\end{equation}
where $\eps(z) = \rhod(z)/\rhog(z)$. Using the settling velocity (\autoref{eq:vsett}) and a vertical diffusivity of $D = \nu$ (i.e. assuming $\St \ll 1$), the solution for the vertical distribution of the dust density becomes
% \begin{equation}
%     \label{eq:rho_d_z}
%     \rhod(z) = \rhog(z) \frac{\rhod(0)}{\rhog(0)} \exp\left\{-\frac{\St_0}{\alpha} \left[\left(\exp\left(\frac{z^2}{2 \hg^2}\right) - 1\right) + \frac{\St_0}{2}\left(\exp\left(\frac{z^2}{\hg^2}\right) -1\right)\right]\right\},
% \end{equation}
\begin{align}
    \label{eq:rho_d_z}
    \rhod(z) & = \rhod(0) \,\exp\left[-\frac{z^2}{2\,\hg^2} - \frac{\St_0}{\alpha}\left(\exp\left(\frac{z^2}{2\,\hg^2}\right)-1\right)\right] \\
             & \simeq \rhod(0) \,\exp\left[-\frac{z^2}{2\,\hd^2}\right]\nonumber
\end{align}
where a vertically isothermal gas density profile was assumed, such that the Stokes number becomes $\St = \St_0 \,\exp(z^2/(2\,\hg^2))$ and the second line assumes $z\ll \hg$. \autoref{eq:rho_d_z} shows that the dust distribution near the mid-plane follows a Gaussian with scale height
\begin{equation}
    \label{eq:hdust}
    \hd = \hg \sqrt{\frac{\alpha}{\St+\alpha}}
\end{equation}
but falls off much faster above one gas scale height. A more general derivation of the transport equation and this equilibrium can be found in \citet{Binkert2023c}.

\autoref{eq:rho_d_z} is the dust distribution in a settling-mixing equilibrium. Similar results can be obtained for pressure maxima in the azimuthal \citep{Birnstiel2013,Lyra2013} and radial direction \citep{Dullemond2018b}.

Radial trapping of dust has been discussed in many works dating back to \citet{Whipple1972}, including \citet{Paardekooper2004}, \citet{Rice2006}, or \citet{Pinilla2012} and many others. It again represents a competition between the radial advective flux (cf. \autoref{eq:nsh1}) and radial diffusion. Radial gradients in the temperature and density make a fully analytical solution difficult and, unlike for the vertical structure, a radial net flux through the pressure bump may exist: a pressure bump that is still receiving dust from the outside and/or lets some dust pass to the inner disk. However, under the assumption of a Gaussian pressure profile and no net radial flux, the steady state surface density again follows a Gaussian with familiar properties
\begin{equation}
    \Sigd(r) \propto \exp\left(-\frac{(r - r_0)^2}{2 \, w_\mathrm{d}^2}\right),
\end{equation}
where $w_\mathrm{d} = w \sqrt{\alpha / (\alpha + \Sc\,\St)}$ and $w$ is the radial standard deviation of the Gaussian gas pressure peak \citep{Dullemond2018b}. For azimuthal pressure peaks, a solution that balances advective and diffusive fluxes can be found as \citep{Birnstiel2013}
\begin{equation}
    \rhod(y) \propto \, \rhog(y) \exp\left(-\frac{\St(y)}{\alpha}\right).
    \label{eq:azimuthal_trap}
\end{equation}
Just like the cases of the radial and vertical directions, the solution again depends on the ratio $\St/\alpha$. This ratio represents the relative strength of advective motion (=drift) and diffusive transport and is therefore by definition the P\'{e}clet number, apart from factors of order unity. In other words: only particles with a Stokes number that is larger than $\alpha$ will experience effective trapping.

The fact that the radial and vertical extent of the dust distribution is sensitive to $\alpha/\St$ means that this ratio can be constrained by observations. This has for example been pioneered in \citet{Pinte2016} for HL Tauri, \citet{Dullemond2018b} for the DSHARP sample of \citet{Andrews2018}. Similar approaches can even constrain asymmetries in radial and vertical direction \citep{Doi2021}, or using disk observations at different wavelengths \citep{Franceschi2023,Doi2023}.

% -----------------------------------------------------------
\subsection{Origins of pressure traps}

As seen in the previous sections, particles have a general tendency to drift towards higher pressures and that diffusion is acting against this tendency. The classical results are that dust sediments to the mid-plane and radially drifts inward \citep{Nakagawa1986}. For localized pressure maxima, this means that dust can be trapped, and the peak concentration is only limited by diffusion. Accumulating large amounts of dust can lead to conditions where planetesimal formation becomes possible \citep{Youdin2005,Johansen2007}. Furthermore, as discussed in \autoref{sec:curtain_opens}, axisymmetric substructure seems to be ubiquitous in planet-forming disks \citep{Andrews2018,Long2018}. Understanding how and where pressure maxima form is therefore a key question in planet formation. A wide variety of mechanisms have been proposed, too many to discuss in depth in this review, which will only summarize the basics in the following (see \citealp{Bae2022} for a review).

Firstly, variations in the disk viscosity $\nu$ can cause pressure maxima due to the fact that the viscous transport speed is proportional to the viscosity. In a steady-state viscous disk, $\dot M_\mathrm{g} \propto \Sigg\,\nu$. A perturbation to the viscosity $\nu$ (or likewise the gas pressure) will therefore cause an inverse effect on the gas surface density \Sigg. This principle is the source of many flavors of pressure bump mechanisms: Dead zones \citep{Gammie1996} are regions in which non-ideal effects (mainly Ohmic resistivity) prevents the MRI from developing. While this simple picture has been complicated by other non-ideal effects (e.g. due to the Hall effect, see \citealp{Bai2015}, or ambipolar diffusion, see \citealp{Bai2011}, \citealp{Gressel2015}), the consensus is, that magneto-hydrodynamic effects will not create the same level of turbulent viscosity everywhere, and such variation will imprint themselves on the pressure profile. A related magneto-hydrodynamic effect is the formation of zonal flows \citep[e.g.][]{Johansen2009,Bai2014,Bethune2016}, self-organizing regions of high magnetic pressure, that can be long-lived enough to trap particles \citep{Dittrich2013}.

The leading hypothesis to form \textit{azimuthal} pressure traps are vortices. Such vortices may be caused by the Rossby Wave Instability \citep{Lovelace1999,Li2000,Chang:Youdin:2023}, by the convective overstability / subcritical baroclinic instability \citep[e.g.][]{Klahr:Hubbard:2014,Lyra2014} or the vertical shear instability \citep{Nelson2013,Richard:Nelson:2016,Latter2018,Manger2020}. Anticyclonic vortices in disks have previously been shown to collect dust particles near their center \citep[e.g.][]{Barge1995,Klahr1997}. More detailed calculations for dust trapping in vortices confirmed the amount of azimuthal concentration in \autoref{eq:azimuthal_trap}, see \citet{Lyra2013}. As predicted by \citet{Wolf2002}, ALMA detected several strongly asymmetric disks \citep[e.g.][]{vanderMarel2013,Casassus2013,Andrews2018,Dong2018} that created lots of interests in vortices as possible sites of planet formation.

\begin{figure}[th!]
    \centering
    \includegraphics[width=.7\textwidth,trim={0 0.31cm 0 0.2cm},clip]{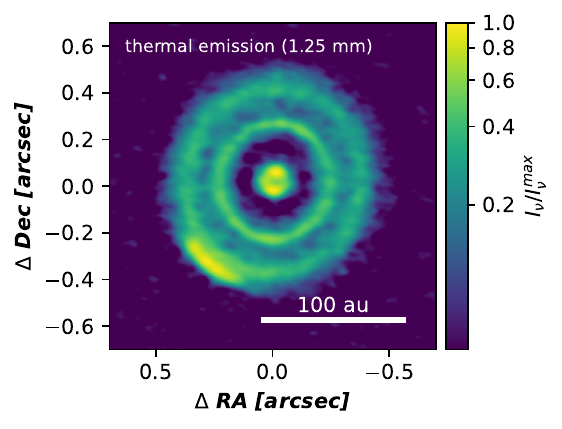}
    \caption{Dust continuum emission at \SI{1.25}{mm} of the disk HD~143006 from the DSHARP Survey \citep{Andrews2018}. The disk within $\sim 0.1\arcsec{}$ appears misaligned \citep{Perez2018}. The substructure was proposed to be caused by a planet between the inner disk and the outer two rings \citep{Perez2018,Ballabio2021}.}
    \label{fig:hd143006}
\end{figure}

Vortices are not the only possible source of asymmetries: massive companions can cause the disk to become eccentric \citep{Kley:Dirksen:2006}. This eccentricity leads to a variation in the orbital speed of the gas with a minimum at apocenter that causes a local accumulation of gas density without particle trapping \citep{Hsieh2012, Ataiee2013}. For high companion-to-star mass ratios above 0.04, this forms a Keplerian-rotating clump with a density contrast of up to 10 \citep{Shi2012,Ragusa2017}. This scenario can not explain asymmetries in disks with multiple rings, such as HD~143006 \citep[][shown in \autoref{fig:hd143006}]{Andrews2018,Huang2018}, or with strong density contrasts, as in  IRS 48 \citep{vanderMarel2013}. \citet{vanderMarel2021} found, that the observed density contrast appears to increase with an observationally estimated Stokes number, as expected from azimuthal trapping \citep{Birnstiel2013}. These observations indicate that asymmetries in the gas may be more abundant than they currently appear, based on asymmetries in the dust emission and that more might be found when moving to longer wavelength and comparable or higher resolutions. These findings make a strong case for azimuthal trapping in vortices, however a small subset of the sources are consistent with asymmetries caused by massive companions, such as HD~142527 \citep{Price2018} where the observed low-mass stellar companion can account for many of the observed features of the circumbinary disk.

Overall, planets are the explanation that currently can explain the majority of substructure: they cause pressure bumps that can trap dust into symmetric rings \citep[e.g.][and many others]{Rice2006,Pinilla2012}. Those pressure bumps can form vortices \citep{Lovelace1999,Li2000,Chang:Youdin:2023} and a single planet can cause multiple rings \citep[e.g.][]{Bae2017,Dong2017}, multiple vortices and spiral arms \citep{LoboGomes2015,Miranda2019}. A planet can also explain the observed perturbations in the rotation velocity around the sub-structure \citep{Teague2018,Izquierdo2022} and the velocity kinks in the channel maps \citep{Pinte2018}. In rare cases, planets have even been imaged \citep{Keppler2018,Muller2018,Benisty2021}. The conclusion is that planets can be responsible for most observed substructures, yet it does not at all mean that all sub-structure is caused by planets, nor that each ring corresponds to one planet. In the future, developments in gas kinematics, and multi-wavelength dust observations as well as more cases of more direct evidence for or against planets will have to shed light on these questions.

% =========================================================
\section{DUST COLLISIONAL EVOLUTION}
\label{sec:coll_evol}

Dust particles in planet forming disks are thought to have initial sizes of $\lesssim\SI{1}{\micro m}$, as grain growth in the low density pre-disk environment is expected to be inefficient \citep{Ormel2009,Bate2022}. The following section will introduce the concepts that determine the collision rates in disks (relative velocities and cross-sections) and the collision outcome, and then discuss the contribution by surface growth (condensation) as well as numerical implementations.

% -----------------------------------------------------------
\subsection{Introduction}

The size evolution of dust can be thought of a general two-body process where kinetics and cross-sections determine collision rates, quite similar to a chemical network. There are however key differences: Firstly, the ``reactants'' are not discrete chemical species, but are samples of a continuous distribution. They can have a range of properties such as mass, composition, or internal structure. For example, many mass combinations of masses $m_1$ and $m_2$ can lead to the same resulting mass $m=m_1 + m_2$. Secondly, the product can be an individual single grain (in the case of perfect sticking), or it can produce a continuum of results, such as a distribution of fragments.

This means that the mathematical representation will be in the form of an integro-differential equation, where the change of a quantity in time (how many particles of a certain type exist) depends on an integral over the size distribution itself. For the mass dimension alone, this can be expressed in a generalized version of the Smoluchowski equation \citep{Smoluchowski1916},

\begin{align}
    \frac{\partial n(m)}{\partial t} = & 
    \int_0^\infty \int_0^{m_1} K(m, m_1, m_2) \, R(m_1, m_2) \,
    n(m_1) \, n(m_2) \mathrm{d}m_1 \mathrm{d}m_2                                            \\
                                       & - n(m) \int_0^\infty R(m, m_1) \, n(m_1)\mathrm{d}m_1,
    \label{eq:smolu}
\end{align}
where we follow the notation of \citet{Stammler2022}. Here, the change of the number density distribution $n(m)$ of particles of mass $m$ has a positive and a negative contribution. The positive contribution is caused by two particles of masses $m_1$ and $m_2$ colliding. This collision happens at a rate $R(m_1, m_2)\,n(m_1)\,n(m_2)$. How much of the colliding mass $m_1 + m_2$ is put into particles of mass $m$ is determined by the quantity $K(m, m_1, m_2)$. For perfect sticking, this would be described by a Dirac delta function, $K(m, m_1, m_2) = \delta(m - m_1 - m_2)$. Some aspects of evolving this equation numerically will be discussed in \autoref{sec:numerics}.

The collision rate $R(m_1, m_2)$ is the product of the collision cross-section and the relative velocity of the colliding particles. It is complicated by two effects: Firstly, one needs to distinguish different outcomes of the collisions, such as perfect sticking, fragmentation, or erosion, and chose the adequate distribution $K(m, m_1, m_2)$. Secondly, the probability of each outcome is velocity dependent. A common choice is to assume an average speed of the particles to compute the collision rate. This might neglect rare but important outcomes because higher- or lower-than-average collision speeds are ignored. It was proposed that a highly unlikely series of events become possible for a very small fraction of the particles, given that the numbers of particles involved in building planets are extremely large (around \SI{e39} micrometer-sized particles are needed to form an Earth mass). This way, few \textit{lucky} particles might be able to continue to grow in an environment where growth would otherwise be hindered \citep{Windmark2012b,Garaud2013}. In case mass is the only particle property of interest and if only mean collision speeds are assumed, the collision rate factor $R_C(m_1, m_2)$ of a given collision outcome $C$ (e.g. sticking) can be simply written as

\begin{equation}
    R_C(m_1, m_2) =
    %\frac{1}{1 + \delta_{m_1, m_2}}
     \sigma_{12} \, \overline{\Delta v}_{12} \, p_C(\overline{\Delta v}_{12}),
\end{equation}
with the collisional cross-section $\sigma_{12}$, the mean relative velocity $\overline{\Delta v}_{12}$, and the probability of the collision outcome $p_C$. These terms will be discussed further below. If the distribution of collision velocities is taken into account, the collision rate factor becomes an integral over this distribution,
\begin{equation}
    R_C(m_1, m_2) = \int_0^\infty
    %\frac{1}{1 + \delta_{m_1, m_2}}
    \sigma_{12} \, \Delta v_{12} \, p_C(\Delta v_{12}) \, p_v(\Delta v_{12}, v_\mathrm{rms})\,\mathrm{d}\Delta v_{12},
\end{equation}
where $p_v(\Delta v_{12}, v_\mathrm{rms})$ is the distribution of relative velocities for a given RMS speed. For the case of a Maxwellian velocity distribution and simple step-functions of the collision probability, analytic solutions can be found \citep[e.g.][]{Stammler2022}.

% -----------------------------------------------------------
\subsection{Relative Velocities}
\label{sec:rel_vel}

For particle collisions to occur, particles need to move relative to each other. Relative motion can be caused by the systematic drift motion of the particles in radial, vertical, or azimuthal direction. As these are size-dependent, particles of different sizes move at different velocities, which results in a relative velocity. Following \citet{Stammler2022} (see also earlier works by \citealp{Weidenschilling1984,Dullemond2005,Tanaka2005,Ormel2007a,Brauer2008}), we can write these contributions as

\begin{equation}
    \begin{split}
        \Delta v_{ij}^\mathrm{rad}  & = \left| v_\mathrm{d,r}(\St_i) - v_\mathrm{d,r}(\St_j) \right|,       \\
        \Delta v_{ij}^\mathrm{azi}  & = \left| v_\mathrm{d,\varphi}(\St_i) - v_\mathrm{d,\varphi}(\St_j) \right|, \\
        \Delta v_{ij}^\mathrm{vert} & = \left|
        h_{\mathrm{d},i} \min\left(\St_i, \frac{1}{2}\right) - 
        h_{\mathrm{d},j} \min\left(\St_j, \frac{1}{2}\right)
        \right| \, \om,
    \end{split}
    \label{eq:relvel}
\end{equation}
for the radial, azimuthal, and vertical relative velocity, respectively, where $h_{\mathrm{d},i}$ is computed as in \autoref{eq:hdust}, for a Stokes number $\St_i$.

In this picture, equal sized particles (or better, particles of the same Stokes number) would not experience collisions. This is circumvented by the fact that the size distribution is never exactly monodisperse, which leads to a dispersion in relative velocities. Furthermore, there are random contributions to the particle velocity which give rise to random velocity for all particles: For very small particles, these are caused by Brownian motion,

\begin{equation}
    \Delta v_{ij}^\mathrm{BM} = \sqrt{\frac{8\, \kB \,T\, (m_i + m_j)}{\pi\,m_i\, m_j}}.
\end{equation}

\begin{figure}[h!]
    \centering
    \includegraphics[width=\textwidth]{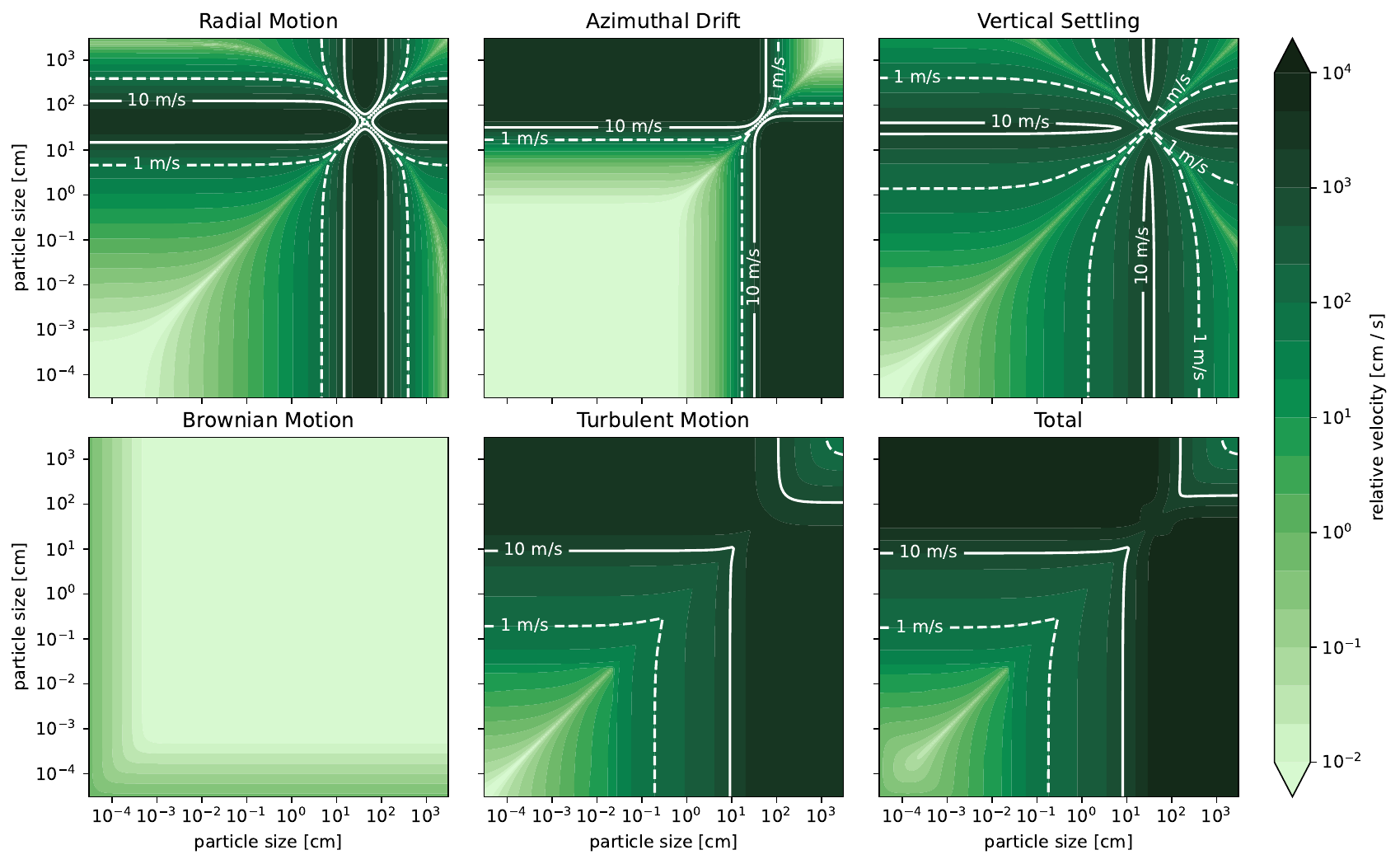}
    \caption{Relative velocity contributions in the standard disk at \SI{1}{au} assuming $\alpha = \SI{5e-4}{}$. The maximum relative velocity is approximately \SI{62}{m.s^{-1}}.}
    \label{fig:relvel}
\end{figure}

Larger particles experience random velocities caused by the drag forces of the turbulent gas. Closed form expressions for turbulent relative motion, based on the model of \citet{Volk1980}, were introduced by \citet{Ormel2007} and are now widely used (however, see \citealp{Pan2014} and following works in that series for more recent results). \autoref{fig:relvel} shows the different contributions to relative velocities as well as the total relative velocities $\Delta v_{ij} = \sqrt{\sum_k \Delta v_{ij,k}^2}$ which is a sum over all contributions. Several trends are visible:
\begin{itemize}
    \item 
          All contributions except Brownian motion tend to increase with the particle size. The maximum is reached at a Stokes number of untity (around \SI{43}{cm} for this example). Only azimuthal drift plateaus at high speeds beyond $\St=1$, other contributions decrease at these sizes.
    \item
          As explained above, only Brownian and turbulent motion have non-zero terms on the diagonal (i.e. equal sized particle collisions).
    \item
          In all cases we find that unequal sized particles (e.g. a large and a smaller grain) tend to collide faster than equal-sized collision pairs.
    \item 
          In this example, $\alpha=\SI{e-3}{}$. As turbulent velocities scale with the sound speed, turbulence is the key driver of relative velocities in the inner disk. In the colder outer disk, radial drift can become the dominant contribution.
    \item
          The transition between Brownian Motion and turbulent relative velocities happens around \SI{1}{\micro m}.
\end{itemize}

\begin{textbox}[h]\section{Turbulent Relative Velocities}
    \label{box:turb_vel}
    Turbulent relative velocities are not as straight-forward to compute as the relative velocity contributions in \autoref{eq:relvel} \citep[see][]{Ormel2007,Pan2014}. While closed form solutions exist \citep{Ormel2007}, is can be helpful and instructive to use approximations that allow order-of-magnitude estimates \citep[see also][]{Birnstiel2011,Powell2019}. Following the approximation of \citet{Ormel2007}, collision speeds between particles of Stokes number \St and monomers of Stokes number $\St_0$ are
    $$ \Delta v^\mathrm{mon}_0 = u_\mathrm{rms} \, \Ret^{1/4}  (\St - \St_0)$$
    if the largest grain has a size below $a_{12} = {2\,\Sigg}/(y_a\,\pi \, \rhos\, \sqrt{\Ret}),$
    where $\Ret = \frac{\alpha \Sigg \, \sigma_{H_2}}{2 \mu \, \mp}$ is the turbulent Reynolds number, $y_a = 1.6$ and $u_\mathrm{rms} = \cs \sqrt{\frac{3}{2} \alpha}$ is the gas turbulent velocity.
    If the larger particle radius $a >= a_{12}$ (intermediate turbulent regime in \citealp{Ormel2007}),
    $$\Delta v^\mathrm{mon}_1 = u_\mathrm{rms} \, \sqrt{\frac{3}{\St + 1 / \St}}$$
    A good approximation therefore is
    $$\Delta v^\mathrm{mon} = u_\mathrm{rms} \, \min\left(\Ret^{1/4}  (\St - \St_0), \sqrt{\frac{3}{\St + 1 / \St}}\right)$$
    while collisions among equal-sized particles can be approximated by
    $$\Delta v^\mathrm{eq} = \sqrt{\frac{2}{3}} \,
        \begin{cases}
            0                     & \text{if } a < a_{12}      \\
            \Delta v^\mathrm{mon} & \text{if } a \geq a_{12}.
        \end{cases}
    $$
    
\end{textbox}

% \begin{figure}
%     \centering
%     \includegraphics[]{figures/rel_vel2}
%     \caption{Relative velocity contributions in the standard disk at \SI{1}{au} assuming $\alpha = \SI{e-3}{}$. Here we compare Brownian motion (equal size collisions), turbulent motion (equal sized, and collisions with monomers) as well as relative drift motion. In the latter, we assume that particles collide with particles of about half their size. $a_\mathrm{BT}$ denotes the transition from Brownian to turbulent relative velocities and $a_{12}$ denotes the transition from fully coupled to intermediate particles as defined in \citet{Ormel2007}.
%     }
%     \label{fig:relvel2}
% \end{figure}

% -----------------------------------------------------------
\subsection{Cross-sections}
\label{sec:cross_section}

The next ingredient to the collision kernel is the cross-section of the respective collision. In the simplest case for spherical particles, this amounts to the geometric cross-section

$$\sigma_{12} = \pi \left(a_1 + a_2\right)^2.$$

For spherical, compact grains (this means uniform porosity or fractal dimension $f_\mathrm{D}=3$), the particle radius is simply related to the mass as $m = 4\pi \rhos \, m^3 /3$. The situation gets more complex if the grains become fractal. This means that the material density varies with radius such that the fractal dimension $f_\mathrm{D}$, defined as $m \propto a^{f_\mathrm{D}}$, is less than 3. In such cases, different definitions are used, such as the characteristic particle radius
$a_\mathrm{c} = \sqrt{\frac{5}{3\,N} \sum_{i=1}^N \vec{r}_i^2}$, where $\vec{r}_i$ is the distance vector from the center of mass of the $N$ monomers that make up the aggregate. This definition is following \citet{Mukai1992} and was used in \citet{Okuzumi2009b} and following works. Earlier works used the area-equivalent radius which is the radius of a circle with the same area as the projected area of the aggregate.

Further modifications of the cross-section can stem from particle charging. In the surfaces of planet-forming disks, dust grains can be positively charged due to photoelectric charging \citep{Pedersen2011,Akimkin2015}. However, in the denser parts of the disk, below the photon-dominated region, grains collide more often with electrons than with ions since the electron velocity is higher than the velocity of the more massive ions. This causes a net negative charge on the grains \citep{Spitzer1941,Okuzumi2009a,Ivlev2016}. The grains will therefore feel a repulsive potential and the collision cross-section needs to be modified with a multiplicative factor,
$$
    f_\mathrm{ch} = \begin{cases}
        1 - \frac{E_\mathrm{el}}{E_\mathrm{kin}} & \text{if } E_\mathrm{el} \leq E_\mathrm{kin} \\
        0                                        & \text{else},
    \end{cases}
$$
where $E_\mathrm{kin} = \frac{1}{2} m_{ij} \Delta v_{ij}^2$ is the kinetic energy of the collision (with $m_{ij}$ being the reduced mass) and $E_\mathrm{el} = Q_i Q_j e^2 / (a_i + a_j)$ is the Coulomb energy barrier at contact between the two grains of charges $Q_i$ and $Q_j$ times the elementary charge $e$.

\begin{marginnote}[]
    \entry{Charging Barrier}{size limit imposed when electrostatic repulsion overcomes the kinetic energy of a collision.}
\end{marginnote}

\citet{Okuzumi2009a} showed that this effect can give rise to a \textit{charging barrier} that would prevent small particles from colliding. It was pointed out that this barrier does not hold for collisions with larger particles that, due to their larger mass and their larger relative velocity possess much more kinetic energy. The growth of larger particles can happen in regions where the charging barrier is not relevant (or affecting only at larger sizes), so possibly in the disk surface, or the inner and outermost parts of the disk \citet{Okuzumi2011}. Transport processes like vertical or radial drift or turbulent diffusion could move those seeds into regions where small grains are prevented from equal size collisions, but could then readily collide with those larger aggregates.

% -----------------------------------------------------------
\subsection{Collisional outcomes}

After having discussed the collision rates and collision speeds, the missing ingredient at this point is the outcome of the collision, often termed the \textit{collision model}. Such a model would need to predict the properties of the resulting particles, depending on the parameters of the collision and the colliding particles. With impact parameter, impact speed, particle composition, particle structure, charges, and more, this is a tremendously large parameter space. Collision models therefore mostly rely on a limited subset of this parameter space, where experiments exist and on interpolating or extrapolating the observed behavior. Experiments can either be done in the laboratory \citep[see, e.g.][]{Blum2008} where analogues of cosmic dust aggregates are produced and collided, mostly under micro-gravity conditions, or the experiments are of numerical nature, where molecular dynamics codes are used to simulate aggregate collisions \citep[e.g.][and many others]{Paszun2009,Wada2009,Seizinger2013,Hasegawa2021}. Both approaches come with the main drawback that we do not know the sizes and properties of the real cosmic dust particles well, the experimentally produced or numerically simulated particles might therefore not represent those well. Furthermore, the microphysics of the collisions is not fully understood, which means that the numeric and experimental results do not always agree \citep{Krijt2015}. This section will summarize the main collisional outcomes and some aspects of current experimental results (for more details see for example \citealp{Blum2018} and references therein).

\subsubsection{Physics of Dust Collisions}

\begin{figure}
    \centering
    \includegraphics[width=\hsize]{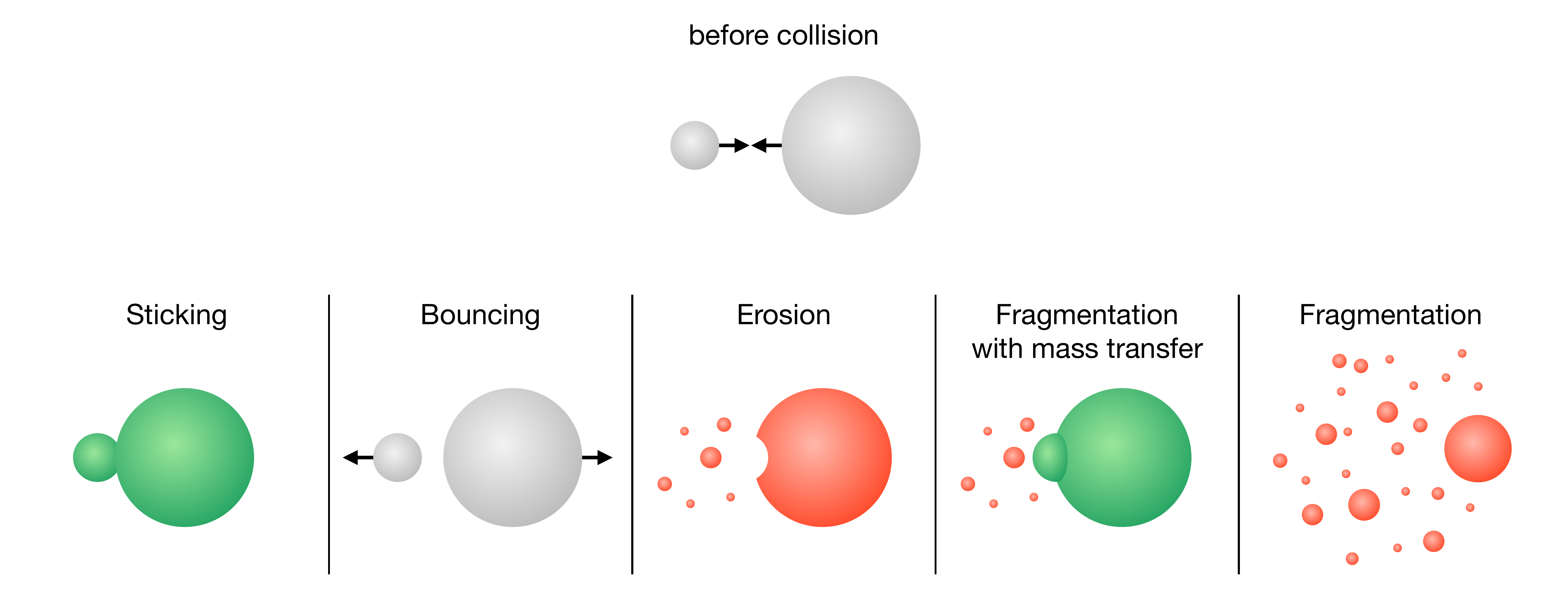}
    \caption{Simplified classification of collisional outcomes. Red and green colors of the collision partners denote net mass loss or gain respectively.}
    \label{fig:coll_outcome}
\end{figure}

The outcome of two colliding dust aggregates is modulated on smallest scales by the surface between the constituent building blocks of the aggregates (see \citealp{Dominik1997} who build on the elastic theory of \citealp{Johnson1971}). Two monomers in contact share a contact area. Deformation of the aggregate requires that these contracts are either broken, or the monomers are rolled, twisted, or slid with respect to each other. \citet{Dominik1997} argue that contact breaking and rolling require the least energy and are therefore the dominant processes. The resulting collisional behavior of aggregates consisting of many monomers typically involve molecular dynamics simulations or is based on laboratory work. The outcome (see \autoref{fig:coll_outcome}) is generally divided into growth-positive collisions (sticking, fragmentation with mass deposition), growth-neutral collisions (aggregates bouncing off each other, usually with some restructuring involved) and growth negative collisions (fragmentation or erosion). \citet{Wada2013} find that the critical collision velocity above which collisions change from growth positive to negative happens at

\begin{equation}
    v_\mathrm{crit} = C \left(\frac{a_0}{\SI{0.1}{\micro m}}\right)^{-5/6},
\end{equation}
where the constant $C$ is around \SI{80}{m.s^{-1}} for water ice and \SI{8}{m.s^{-1}} for silicates. These findings are in good agreement with \citet{Guttler2010} and \citet{Schrapler2011} who used approximately micrometer-sized silicate monomers. The higher erosion threshold velocity for ices was experimentally confirmed by \citet{Gundlach2015}. However, recent laboratory experiments investigated the temperature dependence of the material constants. \citet{Musiolik2019} found that the critical sticking and rolling forces of ice aggregates drop at low temperatures to values comparable to silicates. This would imply that the enhanced stickiness of ices would only apply in a narrow region where the temperature is high enough, but not as high as to sublimate the ices and that the pressure and temperature conditions are crucial in determining the collisional behavior \citep{Gartner2017}.

\subsubsection{Sticking}

As outlined above, collisions that happen below the critical velocity can lead at most to restructuring of the aggregates. Low-speed collisions are therefore mostly of the category sticking depicted in \autoref{fig:coll_outcome}. This outcome can be further subdivided \citep[see][]{Guttler2010} into hit-and-stick at lowest speeds (sticking at first point of contact without relevant structural changes), or, for increasing impact speed, sticking with surface deformations or sticking with penetration.

\subsubsection{Bouncing}

Bouncing is a process that happens mainly at impact speeds beyond sticking and is mainly understood as a collision that does not involve any relevant mass transfer between the colliding bodies, although subtypes of bouncing with mass transfer are discussed in \citet{Guttler2010}. Bouncing involves some amounts of inelastic interaction mainly through compaction around the contact area of the collision. Initially, bouncing was seen in many experimental studies \citep[e.g.][]{Weidling2009,Guttler2010} but could not be reproduced in simulations for similar parameters \citep{Suyama2008,Paszun2009}. Using molecular dynamics simulations, \citet{Wada2011} and \citet{Seizinger2013} showed that the exact method of preparation of the aggregate matters. Their results depend on the \textit{coordination number} and equally on the \textit{filling factor}, while only the latter is easily accessible in laboratory experiments. They conclude that bouncing of particles below \SI{100}{\micro m} is rare, but can happen for compact particles (filling factor above 50\%) and low impact velocities ($< \SI{0.1}{m.s^{-1}}$).

\begin{marginnote}
    \entry{Coordination number}{Average number of contacts of a monomer to its neighbors in an aggregate.}
    \entry{Filling factor, Porosity}{The fraction of the aggregate volume filled with material, as opposed to vacuum/air. Porosity is the volume fraction of vacuum/air.}
\end{marginnote}

\subsubsection{Fragmentation, Erosion, Cratering, \& Abrasion}

Even higher impact velocities tend to lead to mass loss of at least one of the collision partners. The outcome depends on the mass ratio: mass ratios much different from unity tend to lead to destruction of the smaller particle, but often are not energetic enough to fragment the larger particle. Such cases are typically called erosion as depicted in \autoref{fig:coll_outcome}. For less extreme mass ratios, when more mass is excavated, the process is sometimes called cratering \citep{Blum2018} which appears to gradually transition to complete fragmentation (right panel in \autoref{fig:coll_outcome}). Another effect in same family of outcomes is abrasion \citep{Kothe2016,Blum2018} which are small relative mass losses (of the order of a permill in mass) between similar sized aggregates at low speeds.

A special case in this family of outcomes is fragmentation with mass transfer, as observed in experiments of \citet{Wurm2005}, \citet{Teiser2009}, \citet{Kothe2010}, and several following works \citep[see][for a numerical study]{Hasegawa2023}. This effect occurs when a small projectile hits a much larger target at high speed. The impact fragments the projectile, but deposits some mass of the projectile onto the target, similar to a snowball sticking to a wall. As impact speeds tend to increase with particle size (see \autoref{sec:rel_vel}), this provides a potential pathway for continued growth as a larger particle that gets continuously bombarded with smaller grains can slowly but continuously grow as long as it does not experience equal-sized collisions. This possibility will be discussed further in \autoref{sec:towards_planetesimals}.

% -----------------------------------------------------------
\subsection{Condensation/Deposition}

An alternative way of \textit{growing} dust particles is not by mutual collisions (often called coagulation in the context of planet formation), but by deposition from the gas phase to the dust particle surface. This process is often called condensation\footnote{\textit{Condensation}, strictly speaking, is the transition from vapor to liquid state. Under disk conditions the liquid phase usually does not exist, so the vapor is turned to a solid instead. The proper term therefore should be re- or desublimation, or deposition. However, the term \textit{condensation} is still commonly used in the field of planet formation. In the interstellar medium context, this process if often called \textit{accretion}.}, deposition or re-sublimation. The process can be modeled as a flux of atoms or molecules impinging onto the particle surface
$F_\mathrm{gain} = \vth \, n_i / 4$
\citep{Hollenbach2009}, with the vapor number density $n_i$. At the same time, this effect is counteracted by a thermal desorption rate of atoms/molecules leaving the surface, which is given by the Polanyi-Wigner equation \citep[e.g.][]{Minissale2022}, $F_\mathrm{loss} = - N_i\, \nu_i \, \exp( - T_i / T_\mathrm{d})$. Here $N_i$ is the number of desorption sites per area (typically around $\SI{e15}{cm^{-2}}$), and $\nu_i$ is the vibrational frequency of the species $i$ with dimensionless molecular weight $\mu_i$ and a temperature $T_i$ that corresponds to the binding energy $E_i = \kB\,T_i$. The sum of these fluxes determines whether the species sublimates or resublimates and an equilibrium is reached if the vapor pressure reaches the saturation pressure. The desorption rate is complicated by the detailed microphysics of the substrate and by the surface curvature. Often, empirically derived values are used, see \citet{Minissale2022} for details and recommended values of $\nu_i$ and $E_i$.

The sublimation rate above has an exponential temperature dependence. This means that sublimation can be very fast once the temperature rises over the sublimation temperature. The opposite process, resublimation, however, is limited by the collision with the dust surface area. Typically, small particles $\leq\SI{1}{\micro m}$ dominate the available dust surface area. The deposition timescale $t_\mathrm{resub} \simeq \frac{\rhos \, a}{3\,\Sigd}\, \torb$ is then a small fraction of the orbital timescale as long as conditions are far away from equilibrium.

This already leads to the two main issues of particle growth by condensation: Firstly, it is a surface effect, meaning equal amounts of mass are deposited on equal areas. Particle size distributions tend to contain most mass in the largest particles, however most surface area in the smallest particles (as long as $q>3$ in $n(a)\propto a^{-q}$). This means that most vapor is deposited on the smallest grains available instead of significantly growing the largest particles (although surface curvature might affect those rates as well). Secondly, while the timescales can be short, this process quickly runs out of fuel. If vapor is not resupplied, the mass growth stops after a few $t_\mathrm{resub}$. It has been proposed that continuous cycles of sublimation and redeposition can lead to significant growth \citep{Ros2013}, however this neglected the contribution of small silicate particles. \citet{Stammler2017} simulated dust growth and fragmentation together with sublimation and redeposition of carbonmonoxide and found no significant modifications to the overall size distribution, but the CO ice dominated the mass for small particles. A preference for homogeneous over heterogeneous deposition, as proposed by \citet{Ros2019} likely does not change this picture unless the particle size distribution is very narrow.
In any case the required continuous resupply of vapor limits the relevant effects of deposition on dust particles to sublimation fronts.
\begin{marginnote}
    \entry{Sublimation front / line}{Radial distance from the star where the radial temperature profile crosses the sublimation temperature of a given volatile species. Also called ice lines, snow lines, or condensation fronts.}
\end{marginnote}

Even if resublimation might not cause strong effects on the particle growth, the effects of sublimation and resublimation around sublimation fronts, mainly the water snow line, have a wide range of consequences for planet formation. If particles become less sticky upon losing water ice, their reduced drift speed can lead to a traffic jam of solids as shown in \citet{Birnstiel2010} and \citet{Saito2011}. The recondensing water vapor can enhance the solid surface density which can help trigger planetesimal formation \citep[][however, back reactions can hamper these effects, see \citealp{Garate2020}]{Drazkowska2017,Schoonenberg2017,Lichtenberg2021}.

% -----------------------------------------------------------
\subsection{Numerical methods}
\label{sec:numerics}

Analytic treatment of coagulation and fragmentation is generally limited to few simple cases that are of limited applicability apart from validating numerical schemes. They include the constant and linear kernel \citep{Silk1979,Wetherill1990} and Brownian motion \citep{Lai1972}. In some cases a hybrid analytical-numerical scheme can be applied \citep[e.g.][]{Marchand2021}, but most cases need to be treated numerically. Numeric solutions to the coagulation equation are challenging, firstly due to resolution constraints that require large numbers of particles or mass bins and secondly by the limited precision of floating point numbers. A full discussion of the numerics of dust evolution is beyond the scope of this review and only a few references exemplary to the different methods will be discussed briefly. Numerical dust evolution models in protoplanetary disks and beyond can generally be classified into the three categories of mass-grid based, Monte-Carlo based, and approximate models, discussed in the following.

%Growth from \SI{0.1}{\micro m} to \SI{10}{cm} already involves 18 orders of magnitude in mass. Just the linear mass grid itself could already not be stored on a computer which shows that some form of logarithmic grid is needed. For this dynamic range, mass conservation becomes an issue as double precision numbers cannot resolve this (\texttt{1.0 + 1e-16 = 1.0}).

\subsubsection{Mass grid based models}
The particle mass dimension is represented by a grid of size $N$. Dust growth and fragmentation is then based on computing the source and loss terms. Early works using such methods include \citet{Kovetz1969,Weidenschilling1980,Nakagawa1981,Ossenkopf1993,Lee2000}, or \citet{Dullemond2005}. It was shown by \citet{Ohtsuki1990}, that realistic growth can only modeled if more than 7 bins per mass decade are used, otherwise diffusive effects cause excessive underprediction of the growth time scales (while some aspects require much higher resolution, see \citealp{Drazkowska2014a}). Examples of global models that treat dust collisional evolution along with radial transport are \citet{Brauer2008,Birnstiel2010,Okuzumi2012a} or the open source code \texttt{dustpy} \citep{Stammler2022}.

For the case of fragmenting particles, this grid-based approach has computational costs that scale with $N^3$ as there are $N^2$ collisions between the $N$ bins and each collision can affect up to $N$ of the bins through a distribution of fragments. \citet{Rafikov2020} recently showed that for self-similar fragment distributions, the costs can be reduced to $N^2$. Attempts to model dust collisional evolution with fewer numbers of bins were presented in \citet{Liu2017} and \citet{Lombart2021} which look very promising, but have yet to be implemented in multi-dimensional transport or hydrodynamic codes.

The advantage of grid-based integration of the coagulation equation is that it can be quite accurate and, unlike Monte-Carlo methods, it is neither subject to numerical noise nor are some size ranges possibly under-represented. Implicit integration also allows significant speed-up near steady-state solutions. Furthermore, the grid-based treatment of dust mass means that there is no conceptual challenge in implementing it in hydrodynamic grid codes as they are commonly used in the field. The challenges, however, arise due to the numerical diffusion and computational costs. First two-dimensional coagulation simulations coupled to hydrodynamics were shown in \citet{Drazkowska2019}.

A major downside of grid-based models is the challenge of adding additional dimensions beyond the particle mass as this increases both the numeric complexity and the computational costs. While this was done in few cases \citep[e.g.][]{Ossenkopf1993}, recent works used moment approaches to decouple the properties which results in separate coagulation properties, one for the mass and another for the addition properties. This was derived and demonstrated for particle porosity in \citet{Okuzumi2009b} and used to treat the carbonmonoxide abundance in solids in \citet{Stammler2017}.

\subsubsection{Monte-Carlo based models}

Monte-Carlo methods for coagulation have the advantage, that the dimensionality of the model does not increase if additional particle properties are tracked. In all cases, there is a number of particles $N$ that are interacting in a randomized way that sample the true collisional statistics of the model, although larger numbers of particles might be needed to well sample the larger parameter space. Early Monte-Carlo models were presented in \citet{Gillespie1977}, however, when many small particles are joined together to form few larger ones, the number of particles is not conserved and the sampling is reduced. For significant growth, such direct Monte-Carlo methods are therefore usually not suited and some form particle spawning/grouping or representative particles need to be used. Examples include \citet{Ormel2007a,Ormel2008,Zsom2008}. Some models \citep[e.g.][]{Zsom2008} sample mass fractions, which usually means that after substantial growth, no information about the small grain population is retained, once most mass is in the largest bodies. The grouping method of \citet{Ormel2008} can alleviate this issue, however it comes at additional computational complexity. Monte-Carlo methods are a straight-forward way to sample the true statistical variation in the collision outcome, but they also suffer from the numerical noise from limited sampling statistics. \citet{Drazkowska2014a} compared both grid- and Monte-Carlo-based methods for particularly challenging situations of low-numbers of particles. Monte-Carlo based coagulation can also be extended to include particle transport described by a stochastic equation of motion \citep{Ciesla2010b,Ormel2018}, as was done in \citet{Zsom2011b,Drazkowska2013} and others. A coupling to hydrodynamic codes is however challenging as large numbers of particles are required per grid cell in order to well sample the dust distribution. Furthermore, the time step of Monte-Carlo based schemes is bound to the collisional timescale which can be restrictively short. Recently significant advances of representative particle Monte-Carlo methods stem from bucketing methods \citep{Beutel2023}.

\subsubsection{Approximate Models}

Due to the numerical challenges and computational costs of a full treatment of coagulation, simplified or approximate models have  often been preferred. This often allows a speed-up of many order of magnitude, but comes at the cost of now knowing if the approximations are valid. Monodisperse growth is one example, \citep[e.g.][]{Stepinski1997,Kornet2001}, that often allows accurate estimates of the overall particle growth timescale \citep[][]{Dullemond2005,Birnstiel2010}. More accurate are models that assume a given size distribution and approximate the coagulation process with moments \citep[e.g.][]{Garaud2007,Estrada2008,Sato2016}, other models estimate the processes based on the time scales involved \citep[e.g.][]{Ciesla2006,Birnstiel2012,Vorobyov2018}. Modern approaches involve neural networks to speed-up computations and/or to make those approximate treatments more accurate \citep[e.g][]{Pfeil2022}. In the following section, we will follow the treatment of \citet{Birnstiel2012} due to the fact that it is simple, yet well reproduces most parts of global dust evolution models.

% =========================================================
\section{A GLOBAL PICTURE}
\label{sec:global_pic}

Having introduced the transport processes in \autoref{sec:dustdynamics} and the collisional evolution in \autoref{sec:coll_evol}, this section aims at synthesizing a global picture of dust evolution. It is important to realize the difference: simulating the transport of a distribution of small and large grains looks very different from a simulation that evolves this distribution during transport. On global scales dust collisional evolution cannot be neglected and even on small scales, the coagulation timescale can approach the dynamical timescale if the dust-to-gas ratio is increased (see below). In this section we will discuss the relevant time scales, the particle size distribution, and how this knowledge can be used to understand the global behavior of dust in disks on global scales.

\subsection{Time scales}
\label{sec:timescales}

To develop an approximate understanding we turn to the timescales involved. In addition to diffusion and advection time scales, we can derive the collisional timescale by assuming monodisperse growth such that a particle doubles its mass on every collision. The resulting growth rate is then $\dot a = \rhod \Delta v / \rhos$. The actual growth rate thus depends linearly on the relative velocity. As shown in \autoref{fig:relvel}, turbulent velocities are often the largest contribution. If mid-plane conditions, Epstein drag, $\alpha < \St < 1$ are assumed and the relative velocity is approximated with $\Delta v \sim \sqrt{3\,\alpha\,\St}\,\cs$, one finds $\tgrow \sim 1 / (\zed\,\om)$. This approximation works surprisingly well \citep{Birnstiel2012}, however for small particle sizes, low $\alpha$, or different stellar masses, deviations are to be expected because other relative velocities apply \citep[e.g.][]{Powell2019}. To summarize, the important time scales of the system are (with assumptions as in \autoref{eq:simpledrift})

\begin{align}
    \label{eq:ts_grow}
    \tgrow  & \simeq \frac{1}{\zed}\,\frac{1}{\om}                                                        & \text{(growth timescale)}    \\
    \label{eq:ts_drift}
    \tdrift & \simeq \frac{1}{\St\,\gamma} \frac{\Delta x}{r}\, \left(\frac{\hg}{r}\right)^{-2} \frac{1}{\om} & \text{(drift timescale)}     \\
    \label{eq:ts_diff}
    \tdiff  & \simeq \frac{1}{\alpha}\,\left(\frac{\Delta x}{\hg}\right)^2\,\frac{1}{\om}                     & \text{(diffusion timescale)} \\
    \label{eq:ts_sett}
    \tsett  & \simeq \frac{1}{\St\,\om}                                                                       & \text{(settling timescale)}
\end{align}
where the magnitude of the logarithmic pressure gradient is $\gamma = \left|\partial \ln P / \partial \ln r\right|$ which is 2.75 in the standard disk (but larger in the exponential part of the gas density). The timescales correspond to e-folding of the particle size (\autoref{eq:ts_grow}), drifting a length $\Delta x$ (\autoref{eq:ts_drift}), diffusion over length scale $\Delta x$ (\autoref{eq:ts_diff}), and sedimentation towards mid-plane (\autoref{eq:ts_sett}).

\subsection{Growth limits}
\label{sec:growth_limits}

As discussed in \autoref{sec:coll_evol}, dust collisions tend to become growth-neutral or even destructive as the impact speed increases. \autoref{sec:dustdynamics} showed that the collision speeds tend to increase with the particle sizes. As a result, we can expect a maximum size that particles can reach before they stop growing due to bouncing, erosion, or fragmentation. A widely used case is the turbulent fragmentation barrier where the fragmentation threshold velocity \vfrag is equated with an approximate turbulent collision speed which results in a Stokes number of
\begin{equation}
    \St_\mathrm{frag} = \frac{1}{3\,\alpha}\left(\frac{\vfrag}{\cs}\right)^2,
\end{equation}
for Epstein drag and mid-plane conditions, the corresponding particle size is
\begin{equation}
    a_\mathrm{frag} = \frac{2\,\Sigg}{3\pi\,\rhos\,\alpha}\left(\frac{\vfrag}{\cs}\right)^2 \propto \frac{\vfrag^2}{\alpha\,T}.
    \label{eq:afrag}
\end{equation}
For very low values of the turbulence parameter $\alpha$, the maximum turbulent relative velocity $\sim \sqrt{3 \alpha / 2} \, \cs$ can be too small to cause fragmentation. Particles then potentially grow to sizes where the radial drift speed becomes dominant. A drift-induced fragmentation barrier can be derived in the same way using the relative drift speed \citep[see][]{Birnstiel2012},
\begin{equation}
    \St_\mathrm{df} = \frac{2\,\vfrag\,\vk}{\gamma\,\cs^2},
    \label{eq:a_df}
\end{equation}
unless the maximum drift speed is also too slow to cause fragmentation. In that case, the maximum impact speed is reached by azimuthal drift (see \autoref{fig:relvel}), where particles with $\St\gg1$ experience small dust impacting at the full sub-Keplerian speed $\eta\,\vk$, typically several tens of meters per second and likely causing erosion \citep{Krijt2015}. It is worth mentioning that in a pressure maximum, all systematic drift speeds are zero. In this case, only Brownian motion and turbulence remain as source of relative velocity.

\begin{figure}[th]
    \includegraphics[width=0.95\hsize]{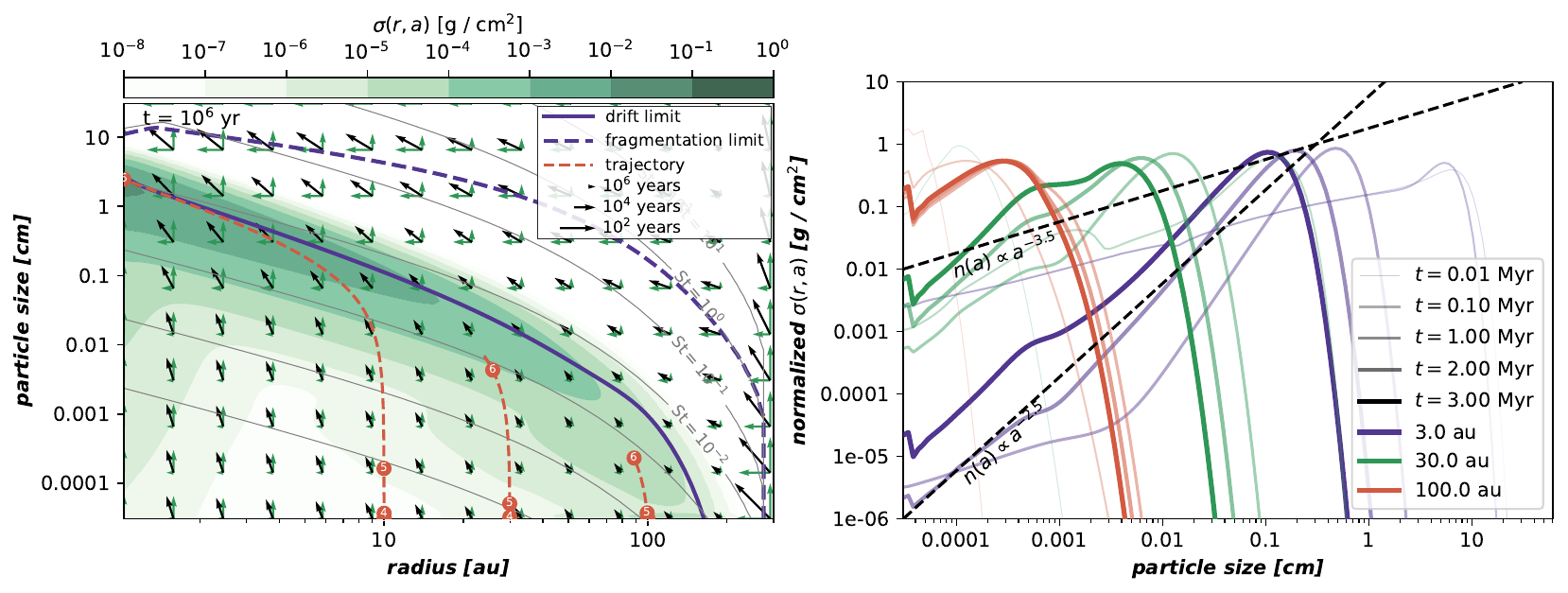}\\
    \includegraphics[width=0.95\hsize]{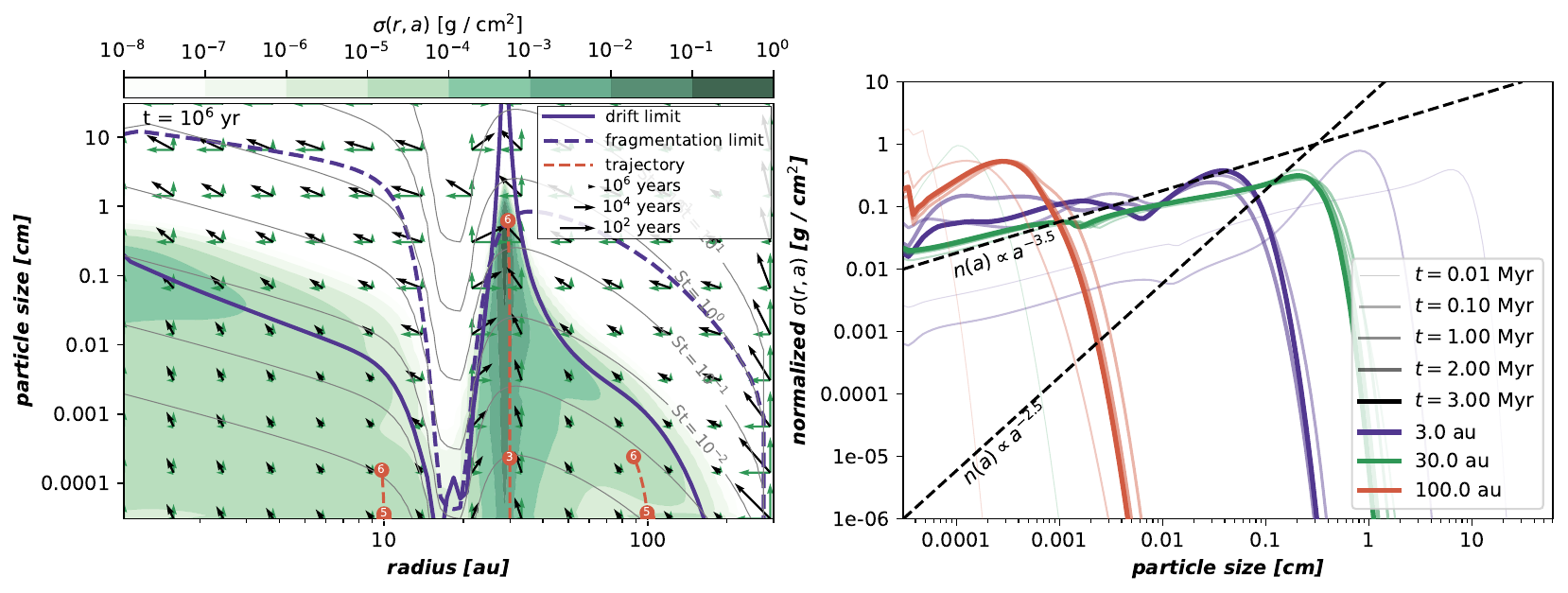}
    \caption{Snapshots of the dust particle size distribution throughout the disk at $\sim$\SI{e6}{years} (left panels) and slices of these distributions at 3, 30, and \SI{100}{au} (right panels). The lower simulation is identical to the upper simulation but includes a deep gap, as expected for a Jupiter-mass planet at \SI{20}{au}. Arrows in the left panels visualize the growth and drift speeds while the red dashed lines indicate the trajectories of a particle under the influences of radial drift and monodisperse growth, where numbers mark the time the particle would need to reach those points, for example \SI{e5}{years} are indicated with the number 5. Purple lines show the size limits imposed by radial drift (solid) and fragmentation (dashed). Gray contours indicate the Stokes number of particles at a given size and radius. The quantity $\sigd(r, a)$ is the dust surface density per decade in size, as discussed in the Box of \autoref{box:sizedistributions}.}
    \label{fig:snapshots}
\end{figure}

For bouncing, a size limit can be derived analogously to \autoref{eq:afrag}, however, the bouncing threshold velocity was found to depend on the particle mass itself. For threshold velocities reported by \citet{Weidling2012} the bouncing barrier has a very weak radial dependency \citep[see for example][Figure 1]{Stammler2023}. In the simulations shown in \autoref{fig:snapshots}, it would be approximately $\SI{2e-3}{cm}\, (r/\SI{10}{au})^{-3/16}$.

The charging barrier \citep[see][and \autoref{sec:cross_section}]{Okuzumi2011} applies to sizes of few micrometers at most, but mixing and drift is thought to alleviate the problem. Furthermore, the particle size measurements (see \autoref{sec:obs_tracers}) appear to show sizes well beyond that barrier, indicating that charging might at most stall, but not prevent particle growth.

The barriers above are all \textit{growth} barriers, meaning that they halt particle growth or make it inefficient. A further particle size barrier that is limiting particle sizes without directly limiting the growth process is the \textit{radial drift barrier} \citep{Birnstiel2012}. It works through transporting larger particles away, faster than particle growth can resupply them. This is not unlike hail stones that cannot remain lifted in the cloud once they reach a certain size. In protoplanetary disks, this limit can be approximated by equating the drift and the growth timescale: with the growth timescale being approximately constant with size this finds the particle size that is removed through drift more efficiently than growth can increase its size. This limit, expressed in terms of Stokes number and in terms of particle sizes, with $\zed = \Sigd/\Sigg$ is
\begin{align}
    \St_\mathrm{dr} & = \frac{\zed}{\gamma} \, \left(\frac{\hg}{r}\right)^{-2}               \\
    a_\mathrm{dr}   & = \frac{2 \, \Sigd}{\pi\,\rhos\, \gamma} \left(\frac{\hg}{r}\right)^{-2}.
\end{align}

\subsection{Particle size distributions}

Another crucial piece of information for any accurate approximation is the particle size distribution. Depending on the physical process, different aspects of the distribution might matter. For chemistry, or vapor deposition the surface area matters, while for the evolution of the dust surface density, knowing the particle sizes that cause the highest dust flux is important. Dust particles in the interstellar medium and also particles in collisional cascades such as in debris disks tend to a size distribution of $n(a)\propto a^{-3.5}$ \citep{Dohnanyi1969,Mathis1977,Tanaka1996}. In planet-forming disks, however, not only fragmentation is at play, but also coagulation. The case of an equilibrium where particles coagulate up to a maximum size at which they break up (fragmentation-coagulation equilibrium) was discussed in \citet{Birnstiel2011} where it was found that the equilibrium size distribution can depend on the size distribution of fragments, unlike in a classical collisional cascade \citep[but see also][]{Gaspar2012,Pan2012}. The resulting steady-state size distributions in fragmentation-coagulation equilibrium can well be understood as a series of analytical approximations, where the exponent of each interval of the size distribution depends on the dominant source of relative velocities in that interval.

\autoref{fig:snapshots} shows size distributions at different positions in dust evolution simulations carried out with \texttt{dustpy} \citep{Stammler2022}. These simulations have a simple collisional model with a sharp threshold velocity that divides sticking and fragmentation (but consider a distribution of impact velocities). If a bouncing barrier were included, this would typically lie between the sticking and the fragmentation regime. Bouncing tends to stall the growth of the largest grains, while small particles might still stick to the largest grains. In such cases, the size distribution can become almost mono-disperse \citep{Zsom2010,Windmark2012a,Windmark2012b,Stammler2023}.

The distribution in \autoref{fig:snapshots} can be divided into two separate types: the fragmentation-limited and the drift-limited distributions. The panels on the left show the size distribution at \SI{1}{Myr} overlaid with the fragmentation and drift size limits. In both cases the mass is dominated by the largest particles. The size exponent is not constant over all sizes, but the fragmentation-limited distributions (the \SI{3}{au} case up to \SI{0.1}{Myr} in both simulations and the \SI{30}{au} case in the lower panels at all times) have a size exponent close to -3.5.

The fragmentation barrier $a_\mathrm{frag}$ depends mostly on variables that are constant or weakly time-dependent and so is the barrier itself. The drift limit, however, has a linear dependence on the dust surface density \Sigd. As drift moves particles inward, the dust surface density decreases with time and therefore also the size limit. 

Dust size distributions that are limited by radial drift tend to be more `top-heavy', where the upper part of the distribution often  follows an exponent of -2.5 (see \autoref{fig:snapshots}). This comes about from the fact that particles are only sticking and growing. Hence, small dust particles are swept-up by the bigger ones and, without replenishment, get depleted. The shape of the top-end of the distribution depends on the level of turbulence: while drift and growth tend to push particles to the drift limit, radial diffusion mixes particles from radially outside the drift limit (i.e. usually smaller sizes) to the inside and, vice versa, particles from inside the drift limit (usually larger particles) outside, hence causing a dispersion in particle sizes. For low levels of turbulence the dispersion can become very narrow. The size distribution in the drift limit for small grains is mostly set by the small grains that are diffused outward from the inner regions of the disk, where fragmentation retains abundant amounts of small fragments \citep{Birnstiel2015}.

\begin{textbox}[h]\section{Particle size distributions}
    \label{box:sizedistributions}
    Particle size distributions are often stated in terms of $n(a) \propto a^{-q}$ (with varying sign conventions of the exponent), where a canonical choice is $q=3.5$ as found in the ISM \citep{Mathis1977} or debris disks and fragmentation cascades \citep{Dohnanyi1969,Tanaka1996}. In this definition, $n(a)$ describes how many particles per infinitesimal size-interval $\mathrm{d}a$ exist per unit of volume. The units are therefore $\SI{}{cm^{-3}.cm^{-1}}$. In the context of protoplanetary disk and planet formation, the available dust mass (not number of particles) is more relevant and due to the disk geometry, the column or surface density is the density of choice. Vertical integration gives a column-number density $N(a) = \int_{-\infty}^{\infty} n(a,z) \mathrm{d}z$, where the vertical dependency of the density was discussed in \autoref{sec:drift_mixing_equi}. The total dust surface density can then be computed as $\Sigd = \int_{0}^{\infty} m(a) \, N(a) \, \mathrm{d}a$. In case the particle size distribution is of interest, not just the total dust surface density, one could define a size distribution in units of surface density $m(a) \, N(a)$, however this would still not well represent the fact that masses and sizes of larger particles are often many orders of magnitude larger than the smallest grains and thus contribute more to the mass integral. It is therefore common practice to define the quantity on a logarithmic scale,
    $\sigd(a) = N(a)\,m(a)\,a,$    
    where the mass integral is understood as being on the logarithmic axis, $\Sigd = \int_0^{\infty} \sigd(a) \, \mathrm{dln}a$. This has the same units as a surface density, \SI{}{g.cm^{-2}}. This quantity, displayed in \autoref{fig:snapshots}, therefore directly visualizes on a logarithmic plot, where most of the mass is located, as it corresponds to the mass per decade in size.
\end{textbox}

\subsection{Global evolution of the dust mass}

Equipped with information on the time scales, the sizes particles can reach, and their distributions, we can draw a global picture of dust evolution. The growth and drift time scales are visualized in \autoref{fig:snapshots} by the arrows, where short arrows indicate long time scales. The growth timescale (\autoref{eq:ts_grow}) shows that particles in the inner disk grow the fastest. Particle growth takes only few decades to proceed. The inward speed of small particles is very slow, except at large radii, where the gas density drops off exponentially. Radial drift is negligible up to a size where the Stokes number (depicted as gray contours in \autoref{fig:snapshots}) exceeds $\alpha$ (within factors of a few, comparing gas and dust radial speeds). Above those sizes, particles start to move significantly relative to the gas. The fact that larger particles drift faster can be seen by the longer inward-pointing arrows in \autoref{fig:snapshots}. This is further visualized by the red dashed lines in \autoref{fig:snapshots} that depict the trajectories of mono-disperse growth, starting at the lower end of the size axis at the time of the snapshot. The numbers denote the time these trajectories take to reach those positions (e.g. 5 denotes \SI{e5}{years}). The trajectories show clearly: 1) how dust growth moves initially vertically (to larger sizes) turns inward (to smaller radii) when the growth and drift time scales become equal at the drift limit and 2) how particle growth and drift happen much faster in the inner disk.

Therefore, radial drift acts akin to an inside-out collapse \citep{Shu1977}: dust at small radii grow quickly, but they also drift inward quickly. On longer and longer time scales, regions further and further out have reached sizes where grains start to drift and supply themselves towards the inner regions.

In the inner parts of the disk, collision speeds are often so high that fragmentation is limiting particle growth. The speed at which the largest grains move is therefore the radial drift speed of particles at the fragmentation size. This speed is not negligible, but considerable smaller than the maximum drift speed at $\St=1$. The drift speed of particles (\autoref{eq:simpledrift}) at the drift limit (for $\St \ll 1$, $\zed\ll 1$) can be written as
\begin{equation}
    v_r(a_\mathrm{drift}) = \zed \, \vk,
    \label{eq:driftlimitspeed}
\end{equation}
which shows that drift quickly removes the first order of magnitude of dust mass. The next order of magnitude in mass is then removed on a 10 times longer timescale (determined by the growth timescale). Globally, the dust-to-gas ratio should therefore be set by the growth timescale of the outermost parts of the disk that still contain a significant dust mass, the \textit{leaky reservoir} as termed by \citet{Garaud2007}. In those regions, the growth and drift time scales become comparable to the age of the disk after about one or two orders of magnitude reduction of the dust-to-gas ratio. Where exactly this reservoir is located is less clear, as the cold outer parts of the disk may be of low surface brightness but could still contain relevant dust masses \citep[e.g.][]{Ilee2022}. As time progresses there will be a smaller and smaller total mass of dust at lower and lower dust-to-gas ratio with longer and longer evolutionary times that supply a smaller and smaller amount of dust flux to the disk inward of it. An outer limit to the dust distribution is imprinted early in the regions where the gas density drops off significantly. There, already the ISM-sized, (sub-)$\mu$m grains would be rapidly drifting towards higher densities. This initial phase of drift can form a sharp outer edge in the dust-to-gas ratio within the first $\sim\SI{0.1}{Myr}$ outside $\sim\SI{100}{au}$ to a few hundred $\SI{}{au}$ \citep{Birnstiel2014}.

\begin{figure}[th]
    \includegraphics[width=0.9\hsize]{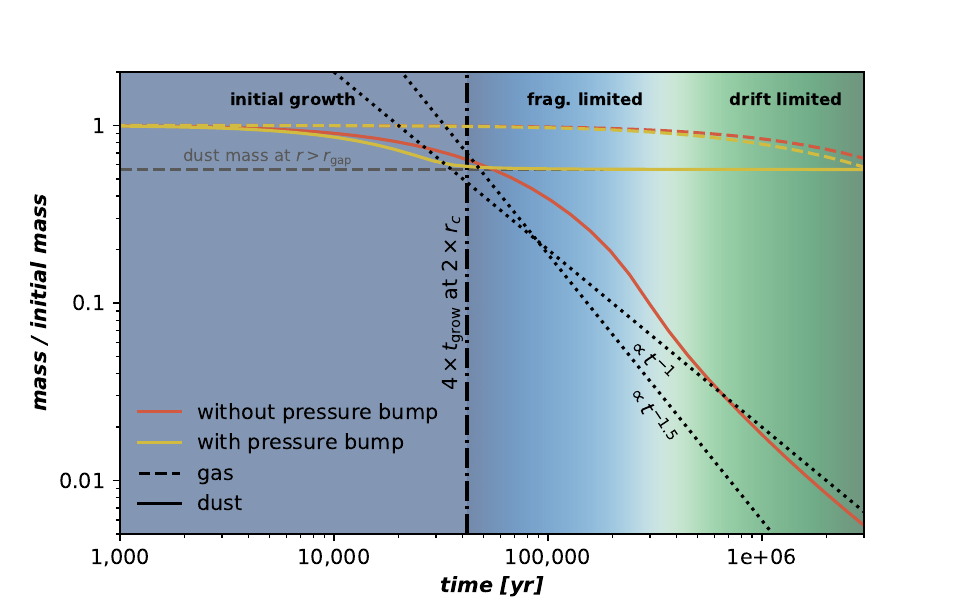}
    \caption{Evolution of the total dust mass (solid) and gas mass (dashed) of the two simulations from \autoref{fig:snapshots}. Orange lines denote the simulation without a pressure trap, yellow lines the simulation with substructure, but no other effects that might reduce the observable dust mass, such as planetesimal formation or pebble accretion. The vertical line shows an appropriate time that particles in the outer disk require to grow and start drifting. Drift-limited growth tends to a time dependency of $\propto t^{-1}$. On timescales longer than the viscous timescale of the disks ($\sim\SI{5}{Myr}$ in this simulation), viscous effects should steepen it towards $\propto t^{-1.5}$.}
    \label{fig:mass_evolution}
\end{figure}

Hence, we can expect that dust in the inner regions of a disk ($r\ll \rc$), if unaffected by substructure, will grow and drain quickly. These regions do contain only a negligible fraction of the total dust mass. Regions further outside contain most of the mass, but their growth lags behind, so during this initial growth phase, we expect an almost constant dust mass in the disk, as seen in \autoref{fig:mass_evolution}. Particles in the outer disk require a few size-doubling time scales to reach the drift limit, so we can estimate this time of constant mass by a few growth time scales (\autoref{eq:ts_grow}) at distances of around 1-2 characteristic radii.

After this initial growth phase (and depending on the collision model), particles can reach large sizes where fragmentation becomes the limiting factor. At this constant size limit (\autoref{eq:afrag}), also the depletion timescale is constant which leads to an initially steep decrease in the dust mass (around \SIrange{0.1}{0.3}{Myr} in \autoref{fig:mass_evolution}). This decrease in dust mass (and dust-to-gas ratio) moves the drift limit towards smaller particle sizes until it becomes the limiting growth barrier. This is seen in the top-left panel of \autoref{fig:snapshots}, where fragmentation only happens inside of \SI{1}{au} while the entire outer disk follows the drift limit. Under these conditions, we can expect the majority of the dust mass to drain with a speed of \autoref{eq:driftlimitspeed}, which therefore tends to a $t^{-1}$ dependency, as shown in \autoref{fig:mass_evolution} after about \SI{0.5}{Myr}. If this timescale approaches the viscous timescale, it will tend towards $t^{-1.5}$ due to the additional mass loss through viscous transport.

If significant sub-structure is present, this process can look drastically different, as shown in the lower panels in \autoref{fig:snapshots} and yellow lines in \autoref{fig:mass_evolution}: the displayed simulation is identical apart from the fact that a modification in the gas viscosity imposes a gap (using the gap profile of \citealp{Duffell2020} for Jupiter mass planet). The gap at \SI{20}{au} causes a pressure maximum close to \SI{30}{au}. With the pressure gradient approaching zero towards the maximum, radial drift as well as radial and azimuthal relative velocities vanish. This means that only the fragmentation barrier (or a bouncing barrier if included in the collision model) remains as growth barrier as long as the gas turbulent velocity is high enough for fragmenting collisions. The size limits in the lower left panel of \autoref{fig:snapshots} indicate that particles in the pressure maximum are fragmentation limited, while the entire rest of the disk is already in the drift limit. The radial directions of the arrows show that this pressure maximum moves the particles between 20 and \SI{30}{au} radially outward towards the pressure bump while the dust outside the trap and inward of the gap keeps drifting radially inward. The dust inside \SI{20}{au} is therefore drained, while practically all dust outside \SI{20}{au} is steadily transported into the pressure trap. \autoref{fig:mass_evolution} shows that in this simulation, the mass evolution levels off at approximately the initial dust mass outside the gap.

\subsection{Global evolution of the dust surface density}

The size distributions discussed previously were shown to be \textit{top-heavy}, i.e. contain most total mass in the largest grains. At the same time, the largest grains tend to have the highest drift speeds. Therefore, the total dust flux (sometimes called the pebble accretion rate) is dominated by the upper end of the size distribution. This upper end can be simply approximated: initially all particles are small, but they grow at a speed of $\dot a \simeq \rhod \, \Delta v / (2\, \rhos)$ (see \autoref{sec:timescales}), which is often dominated by turbulent relative velocities and then becomes $\dot a \simeq a\, \zed \, \om / 2$ (see \autoref{eq:ts_grow}). After a few local growth time scales, the particles reach one of the size limits described in \autoref{sec:growth_limits}, whichever is more constraining. At this point, the dust mass flux $\dot M_\mathrm{d} = 2\,\pi\,r\,\bar v_{d,r} \,\Sigd$, is largely set by the mass and drift speed of the largest grains, or stated differently, the mass averaged velocity $\bar v_{d,r} = (\int_0^\infty v(a)\,\sigd\,\mathrm{dln}a) / \Sigd$ will be within a factor of order unity of the drift speed of the largest grains. \citet{Birnstiel2012} therefore proposed a two-population model where the entire dust distribution is represented by only two populations of dust: small dust that effectively follows the gas radial speed (and contains an often negligible fraction of the mass) and a population of large dust that moves with the drift speed of the largest grains, multiplied by a fudge factor that is calibrated by numerical simulations.

If we ignore this order-of-unity factor and the small contribution of small dust grains, we can (slightly over-)estimate the total dust flux as $\dot M_d = 2\pi\,r\, \bar v_{d,r}(\amax) \, \Sigd$. The previous subsection showed that the outermost regions of the disk retain their dust the longest, so they provide a slowly decreasing reservoir of dust mass and set the value of $\dot M$. The regions inward will mostly just transport this dust further inside with the velocity given by the largest grains. This allows solving the mass flux equation for the surface density,
\begin{equation}
    \label{eq:sigmad_steadystate}
    \Sigd \simeq \frac{\dot M_d}{2\pi\,r\,\bar v_{d,r}(\amax)},
\end{equation}
where we can now insert which ever size limit sets \amax. As shown in \citet{Birnstiel2012}, this would predict a dust surface density of $\Sigd\propto r^{-3/4}$ for a gas density of $\Sigg \propto r^{-1}$ if drift is the limiting growth barrier. For the fragmentation barrier (and constant fragmentation velocity and $\alpha$), this would always result in a profile of $\Sigd \propto r^{-3/2}$. \autoref{eq:sigmad_steadystate} also shows that variation in the fragmentation velocity cause the maximum particle size to vary (see \autoref{eq:afrag}) and opposite variations in the dust surface density. As shown in \citet{Birnstiel2010}, lower fragmentation speeds inside the snow-line would cause an increase of the surface density. The same mechanism was applied in \citet{Okuzumi2016} where sintering was proposed to lower the fragmentation velocity in the regions outside sublimation fronts and this way causing local enhancements in the dust surface density.

% =========================================================
\section{THE PATH AHEAD}
\label{sec:path_ahead}

The previous sections drew a broad picture of the processes and timescales involved in particle growth and transport, the earliest stages of forming planets. The next sections will look beyond those processes towards testing the predictions made via observational tracers (\autoref{sec:obs_tracers}), towards growth beyond the growth barriers (\autoref{sec:towards_planetesimals}) and then conclude with some future directions of the field.

% -----------------------------------------------------------
\subsection{Observational Tracers}
\label{sec:obs_tracers}

In \autoref{sec:curtain_opens}, some observational trends have already been introduced: Within the last decade, it became possible to observe planet-forming disks at high angular resolution and sensitivity, thanks in particular to the ALMA Observatory and the SPHERE instrument at the VLT. ALMA covers a wavelength range from a fraction of a millimeter to a few millimeters and both ALMA and SPHERE have capabilities to observe polarized light. These capabilities enable not just imaging the disks, but also to develop observational probes of the underlying physics. In the following, the most important observational tracers and trends and their impact on our picture of early planet formation will be discussed. For more details, the reader may refer to the review in this series by \citet{Andrews2020}.

\begin{marginnote}[]
    \entry{Population properties}{Observable properties of the stars or disks that can be measured for large numbers of sources, including disk integrated fluxes, spectral indices, disk sizes, dust/gas masses, or stellar masses, ages, and accretion rates.}
    \entry{Effective radius}{The radius encompassing a defined fraction of the total flux, often 68\% or 99\%, depending on application.}
    \entry{Albedo}{The (single-scattering) albedo $\omega = \bar \kappa_{\nu,\mathrm{sca}} / (\bar\kappa_{\nu,abs} + \bar\kappa_{\nu,sca}$ is the probability of a photon being scattered instead of being absorbed when interacting with a dust grain.}
\end{marginnote}

\begin{figure}[th]
    \centering
    \includegraphics[width=\textwidth]{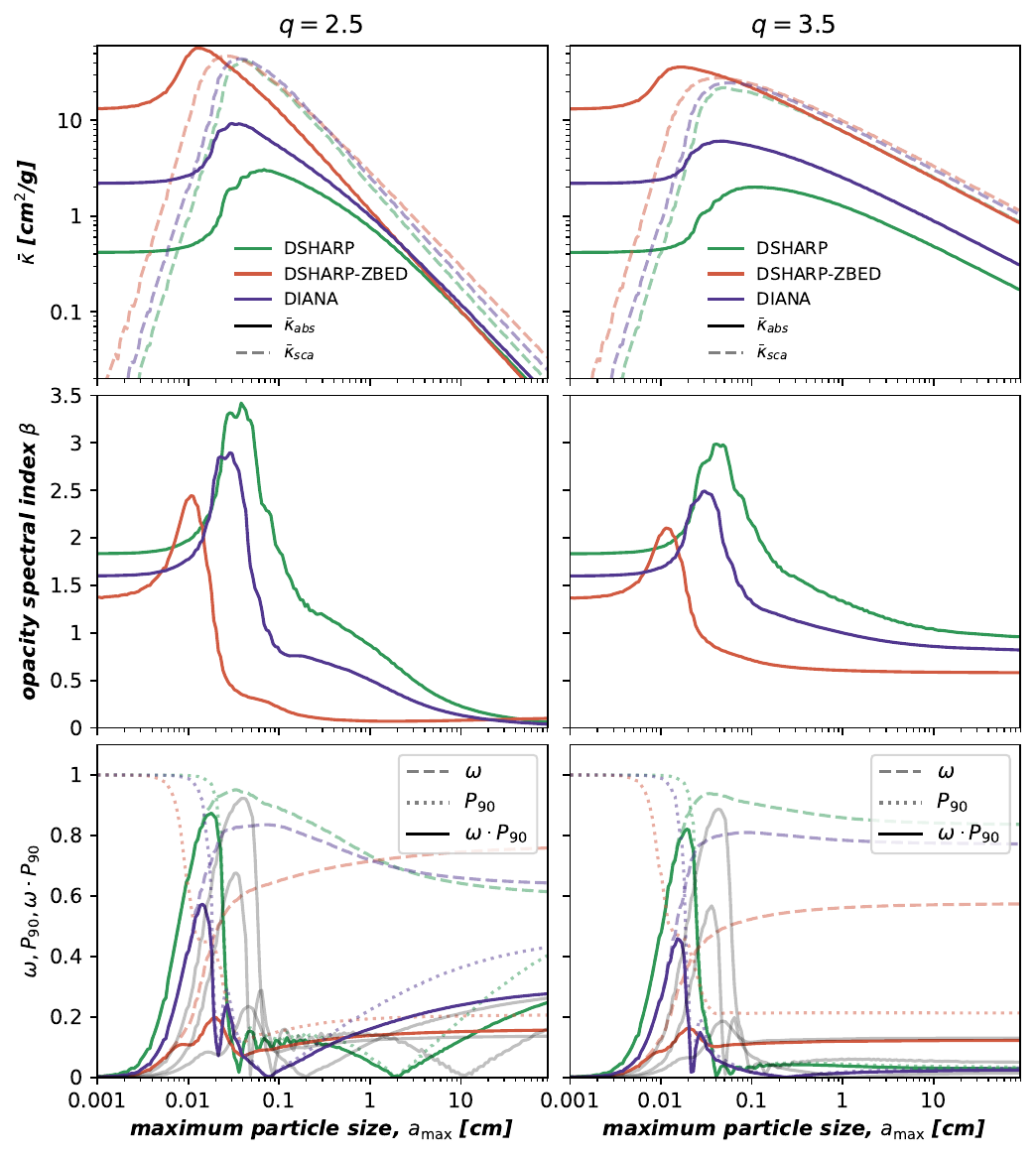}
    \caption{Mass absorption and scattering opacity (top row, at \SI{1.25}{mm}), spectral indices (middle row, measured between 1.25 and \SI{3}{mm}), and scattering properties of dust particles as function of maximum particle size \amax, averaged over a size distribution with size exponent of $q=2.5$ (left) and $q=3.5$ (right). Lines correspond to default DSHARP (green), DSHARP with different carbonaceous material and computational method (orange, labeled DSHARP-ZBED) and default DIANA (purple). Solid and dashed lines in the top row correspond to absorption and scattering, respectively. The bottom panels show the albedo $\omega$ (dashed), the probability of perpendicular scattering $P_{90}$ (dotted), and the product of both (solid) at \SI{1.25}{mm}. Gray lines show how these values are shifted for a wavelength of \SI{3}{mm}.}
    \label{fig:opacity}
\end{figure}

\subsubsection{Dust opacities and dust spectral properties}

In the context of dust continuum emission, the spectral index is defined as the exponent of the spectral energy distribution in the (sub-)millimeter range, $F_\nu \propto \nu^\alphamm$, sometimes given with a specific wavelength range over which it was measured, such as $\alpha_\mathrm{1-3mm}$. This should not be confused with $\alpha$ (without subscript), that denotes the turbulence parameter in this review. In a simple slab model, neglecting the effects of scattering or background, the emitted intensity is
\begin{equation}
    I_\nu = B_\nu (1 - \exp(-\tau_\nu)) = B_\nu\,\times\begin{cases}
        1 & \text{for } \tau_\nu \gg 1\\
        \tau_\nu & \text{for } \tau_\nu \ll 1,
    \end{cases}
    \label{eq:inu}
\end{equation}
where $B_\nu$ and $\tau_\nu$ are the Planck function and the absorption optical depth along the line of sight. The latter is often well approximated by $\tau_\nu = \Sigd\,\bar \kappa_\nu / \cos i$, where $\bar \kappa_\nu$ is the opacity (absorption cross-section per gram of dust, units of $\SI{}{cm^2.g^{-1}}$), averaged over the dust size distribution ($\bar \kappa_\nu = \int \sigd(a) \, \kappa_\nu(a) \mathrm{dln}a / \Sigd$), and $i$ the inclination.

If the opacity follows at least locally a power-law, $\bar \kappa_\nu \propto \nu^\beta$, the frequency dependency of $I_\nu$ in the Rayleigh-Jeans limit becomes $\alphamm = 2 + \beta$ for optically thin emission and $\alphamm = 2$ for optically thick emission (where the number 2 might be smaller if the emission is not deep enough in the Rayleigh-Jeans limit). \autoref{fig:opacity} shows the optical properties of two particle size distributions (with size exponents $q=2.5$ and $q=3.5$) of different materials and employing different computational methods. DSHARP denotes the standard choice of \citet{Birnstiel2018} (mostly backward-compatible with \citealp{Dalessio2001}), DSHARP-ZBED uses the same mass fractions, but uses different optical constants for the carbonaceous materials, \citet[sample `BE']{Zubko1996}, instead of the ones by \citet{Henning1996} and furthermore uses DHS (\citealp{Min2005}, with their parameter $f=0.8$) to compute the opacities, instead of classical Mie theory. DIANA denotes the standard opacities of \citet{Woitke2016}. This shows that the choice of carbonaceous material, and to some extent also the computational method, can strongly affect the shape and strength of the absorption opacity with very limited effect on the scattering opacity (see top row in \autoref{fig:opacity}). The particle size distribution exponent $q$ affects particularly the opacity spectral index $\beta$ for large maximum particle sizes (comparing the left and right columns of \autoref{fig:opacity}). For the chosen opacities, only the top-heavy particle size distribution with $q=2.5$ (which resembles the radial drift limited simulations shown in \autoref{fig:snapshots}) reaches very low values of $\beta$. For a given (uncertain) assumption of the opacities, measurements of the spectral index can thus in principle be used to constrain the maximum particle sizes in the sub-millimeter to centimeter range \citep[e.g]{Ricci2010,Isella2010,Perez2012}. This approach is further complicated by the unknown optical depth, which can mimic a low dust opacity spectral index via high optical depth as both lead to $\alphamm \sim 2$ \citep{Ricci2012}. The measured low values of $\alphamm$ (see next section) are indicative of significant grain growth, high optical depth, or likely a combination of both.

If scattering is considered \citep{Miyake1993,Birnstiel2018}, the resulting intensity in the optically thin region is increased (a photon takes a longer path and appears more thermalized with the medium), while the optically thick emission is reduced (see Figure~7 of \citealp{Miotello2023}, or also \citealp{Zhu2019}). The amount of reduction in the latter case depends on the albedo $\omega = \bar \kappa_{\nu,\mathrm{sca}} / (\bar\kappa_{\nu,\mathrm{abs}} + \bar\kappa_{\nu,\mathrm{sca}})$. At wavelengths, where the albedo is strongly wavelength-dependent (see bottom row in \autoref{fig:opacity}, near \SI{0.2}{mm}), this might lead to spectral indices even below $\alphamm =2$ \citep{Liu2019}.

For increasing maximum particle size around $\amax \sim \lambda / (2\pi)$ the probability of scattering ($\omega_\nu$) increase while at the same time scattering becomes more directed into the forward direction. Perpendicular scattering produces the most polarized intensity, but the probability of perpendicular scattering, $P_{90}$, decreases with \amax. The combination of scattering becoming more likely, but less isotropic leads to a sharp peak in the degree of polarized intensity \citep{Kataoka2015}, as shown in the bottom row of \autoref{fig:opacity}. This polarization by self-scattered dust emission at different wavelength allows setting tight constraints on the dust grain size at around \SIrange{0.1}{0.2}{mm} \citep{Kataoka2016,Kataoka2017}, although this picture is complicated by other mechanisms that can cause polarization at radio wavelengths \citep{Kataoka2017,Kataoka2019}.

At the same time, multi-wavelength analysis of the continuum emission can be used to constrain the maximum particle size, dust surface density, and temperature profiles of disks, \citep[e.g.][]{Carrasco-Gonzalez2019,Ueda2020,Sierra2021,Guidi2022} often finding slightly larger particle sizes around \SI{1}{mm}. This inconsistency between measuring particle sizes with (1) the polarization signal and (2) the continuum spectral index might be solved by the contributions of optically thick emission, as proposed in \citet{Lin2020b} for HD~163296. Modeling the rings in this disk, \citet{Doi2023} computed much larger values, however, they also show how much the absolute values of \amax depend on the choice of opacity. In the future, it will be instructive to bring different methods of measuring particle sizes in agreement when applied to the same disk. \citet{Doi2023} also derive constraints on the turbulence parameter, finding relatively high values of $\alpha \sim \SI{6e-2}{}$ in the vertical direction, while similar approaches for other disks indicate extreme levels of settling, consistent with $\alpha < 10^{-4}$ \citep{Villenave2022}. This shows the diversity of disk physical properties which will require studying not only individual sources, but also population statistics of disks in large numbers, as discussed in the next section.

\subsubsection{Population properties}

\begin{figure}[h!]
    \centering
    \includegraphics[width=\textwidth]{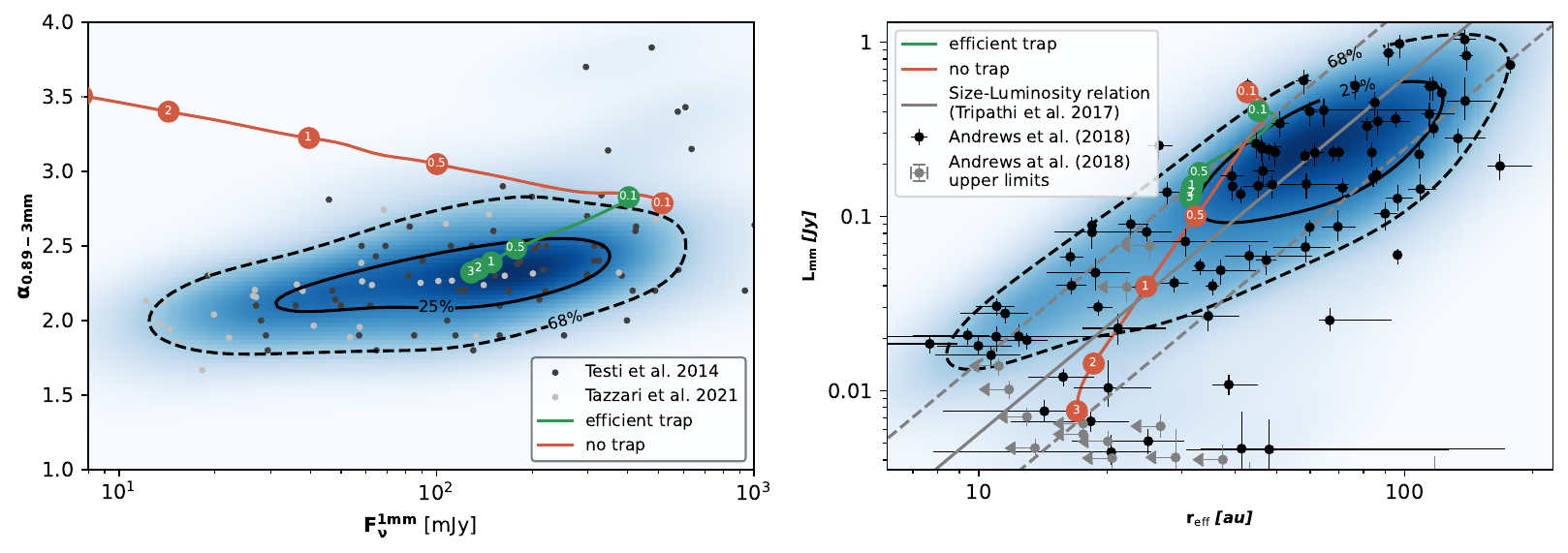}
    \caption{Left: Distribution of millimeter fluxes and spectral indices compared to simulations of dust evolution simulations. Black and gray dots are the samples of \citet{Testi2014} and \citet{Tazzari2021b}, respectively. Their combined kernel density estimate distribution is shown in color scale where the solid/dashed black contour encompasses 25 and 68\%. Right:  Size-Luminosity correlation (gray lines) of \citet{Tripathi2017} and \citet{Andrews2018a} and the samples from which the correlation was derived (points and kernel density estimate). In both panels, colored lines depict the evolutionary tracks of the simulations shown in \autoref{fig:snapshots} and described in the text box ``the standard disk'', where numbers denote the simulation time in millions of years. The green line is a simulation with an efficient dust trap, the orange a simulation without a trap.}
    \label{fig:populations}
\end{figure}

Population properties such as integrated fluxes or radii were accessible already pre-ALMA, but population studies benefitted strongly from the high sensitivity of ALMA. Studies from \citet[for the Taurus star forming region]{Andrews2013}, \citet[for Lupus]{Ansdell2016}, \citet[for Upper Scorpius]{Barenfeld2016}, \citet[for Chamaeleon I]{Pascucci2016}, \citet[$\sigma$ Orionis]{Ansdell2018}, and \citet[of Class 0/I sources in Perseus]{Tychoniec2018} and many others brought the numbers of observed disks close to 1000 (see \citealp{Miotello2023} and \citealp{Manara2023} for recent reviews). These studies allow moving beyond studies of individual disks and enable probing general correlations or age trends that any theory of disk evolution needs to be able to explain. \citet{Tripathi2017} for example discovered a disk-size-flux relation, often called size-luminosity-relation (SLR, see also \citealp{Andrews2018a}) in which the integrated dust continuum flux scales very closely with $\reff^2$. This is surprising since disks do not display a radially constant surface brightness. \citet{Rosotti2019} and \citet{Zormpas2022} showed that drift-limited particle growth could naturally explain this trend. However, this seems at odds with the fact that significant substructure is deemed necessary to explain (1) high-resolution images \citep[e.g.][]{Andrews2018,Dullemond2018b} as well as (2) the generally low spectral indices of disks, shown in the left panel of \autoref{fig:populations}. For comparison, two evolutionary tracks of two simulations are shown overlaid on an observed population of disks \citep[from][]{Testi2014,Tazzari2021b,Andrews2018a}. The green line depicts a simulation with a strong pressure trap (as caused by a Jupiter-mass planet, based on the profile of \citealp{Duffell2020}, corresponding to the lower row of \autoref{fig:snapshots}). The orange line shows the identical simulation but without a pressure trap (corresponding to the upper row in \autoref{fig:snapshots}). Since radial drift preferentially removes the large grains, the spectral index in the latter case increases as the dust mass (and therefore also the millimeter flux) decreases. It can be seen that the simulation with pressure trap is broadly consistent with the observed values. However, large numbers of simulations are required to draw firm conclusions and lead to the expectation that sub-structured disks would follow a shallower size-luminosity trend as observed \citep{Zormpas2022}.

It is remarkable that none of the observed disks shows high spectral indices towards lower-flux-ranges. This is indicating that dust grains remain large and/or optically thick throughout their later evolution, at least at the flux levels where spectral indices are currently available. The very low spectral indices below $\alphamm=2$ are partially attributed to errors, but some can be caused by the effects of scattering \citep[see][and the discussion above]{Zhu2019,Liu2019}.

While population synthesis of disks is still in the infancy \citep[see][for some recent work]{Rosotti2019b,Somigliana2020,Zagaria2022,Zormpas2022}, future efforts are needed to find the initial conditions and the disk evoluationary processes that can explain the large ranges of population properties available today (gas and dust mass constraints, multi-wavelength-dust and gas disk sizes, spectral indices, accretion rates, and more). A more in-depth discussion of disk populations and the measurements of disk fundamental properties can be found in \citet{Manara2023} and \citet{Miotello2023}.

% -----------------------------------------------------------
\subsection{The growth towards planetesimals}
\label{sec:towards_planetesimals}

Given the size limits set by bouncing, fragmentation, erosion, or radial drift, it might seem as growth towards gravitationally bound bodies, called \textit{planetesimals} is impossible. There are however proposed pathways of growth beyond the size barriers.

Firstly, \emph{lucky particles} were proposed by \citet{Windmark2012a}, \citet{Windmark2012b}, and \citet{Garaud2007}. As discussed in \autoref{sec:coll_evol}, this relies on the fact that small particles colliding at tens of meters per second may deposit mass onto much larger target particles \citep[e.g.][]{Teiser2009}, so a single larger particle might be able to grow in an environment where all other particles remain limited to small sizes by a barrier. Producing such a single larger particle might be possible given the randomness in the collision speed (or other collision parameters), such that a particle could happen to be always be colliding with particularly beneficial outcome, giving rise to the term \emph{lucky} particles. Such a scenario is very improbable, but given the large numbers of particles, conceivable. However, there are two main caveats of this idea. Firstly, the radial drift barrier cannot be circumvented by lucky collisions as it is not a collisional, but a dynamical barrier, it would have to be prevented by a pressure trap. Secondly, even if the large particle is able to grow, its growth timescale would be very long, too long to efficiently produce the planetessimals needed for the formation of planets \citep[e.g.][]{Estrada2016}.

Secondly, \emph{fluffy growth} was proposed as a way to grow quickly enough to bypass the radial drift barrier. When including porosity evolution in models of particle growth and transport \citep[e.g.][]{Okuzumi2009b}, it was found that large particles in the inner disk can grow through the drift barrier \citep{Okuzumi2012a}. This is enabled by the fact that particles in the inner disk, if gas densities are high enough, can experience Stokes drag instead of the Epstein drag (see \autoref{sec:dragforces}). In that case, the growth timescale is not independent of size (as in \autoref{eq:ts_grow}), but inversely proportional to the particle size. Therefore, particles with a low fractal dimension and consequently large cross-section can reach very short growth time scales \citep{Okuzumi2012a, Kataoka2014}. This, however, requires that the particles do not fragment, even at tens of meters per second \citep[see][]{Krijt2015}. Furthermore, highly porous particles currently do not seem to be able to reproduce observed disk populations \citep{Zormpas2022} due to their low opacity \citep{Kataoka2014}. Still, more work is required to draw firm conclusions about particle porosity.

\begin{figure}[th]
    \includegraphics[width=\hsize]{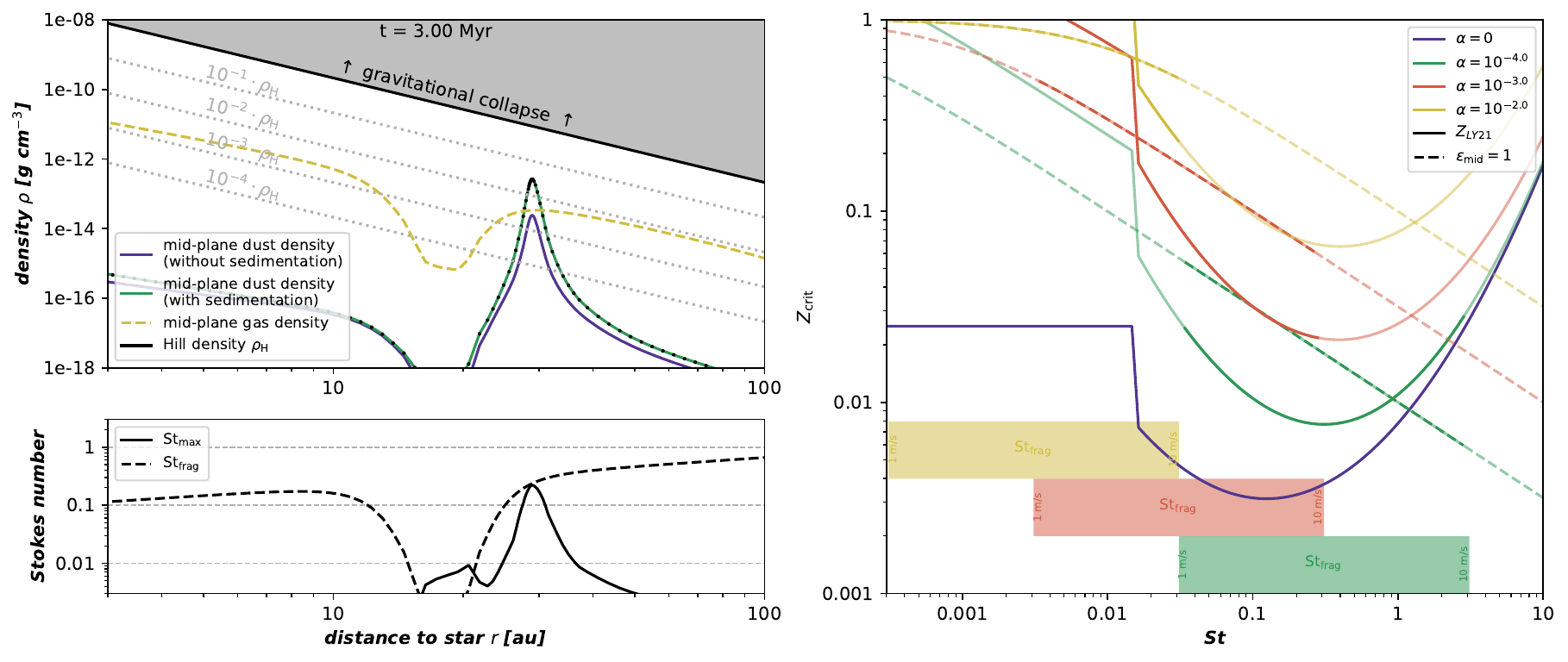}\\
    \caption{Left: Simulation of dust accumulation in a pressure bump based on the simulation shown in \autoref{fig:snapshots} at \SI{3}{Myr}. Neither dust-back-reaction nor planetesimal formation are included, thus the dust-to-gas ratio is exceeding unity in the pressure bump. The bottom left panel shows that dust particles are large and limited by fragmentation in the bump, in contrast to the rest of the disk. Right: the lower limits of the metallicity $Z$  required for strong-clumping for different levels of turbulence. The boxes indicate the Stokes numbers that dust evolution models can reach in the fragmentation limit.}
    \label{fig:hill}
\end{figure}

The currently favored scenario of planetesimal formation is \textit{gravitational collapse of a dust cloud}, similar to the original works of \citet{Safronov1969} and \citet{Goldreich1973}. The challenge in this scenario are the many orders of magnitude difference between the initial dust volume density and the density that is required for the dust cloud to be gravitationally bound, which happens at around the Hill density $\rho_\mathrm{H} \simeq 9 M_\star / (4\pi\, r^3)$. At this point, the gravitational collapse can proceed until material densities of order unity in CGS units are reached, unless effects on smaller scales intervene \citep{Klahr2021}. \autoref{fig:hill} displays the dust, gas and Hill densities in the mid-plane of a disk at \SI{3}{Myr}. The dust, at the canonical dust-to-gas ratio of 1\% needs to be accumulated by more than four orders of magnitude. As discussed in \autoref{sec:drift_mixing_equi}, dust is sedimenting towards higher pressure, so in a ring of high pressure, dust is accumulated in the radial and vertical direction, as seen at around \SI{20}{au} in \autoref{fig:hill}. This can raise the dust-to-gas ratio towards unity. At this point, the dust mass and momentum are not negligible compared to the gas and therefore, the dust starts to strongly affect the gas dynamics \citep[e.g.][]{Fu2014,Taki2016,Onishi2017,Dipierro2018,Raettig2021}. Initially, it was thought that those dynamical affects are detrimental to planet formation \citep{Cuzzi1993} as the dust-gas shear motion would trigger turbulence which through turbulent mixing inhibits further concentration of the dust. The other face of turbulence is, however, that it does not act as a homogeneous diffusion on all scales, but that it can also form localized pockets of high and low densities. Under the right conditions (see below), one effect coming into play is the \textit{Streaming Instability} (SI) \citep{Youdin2005,Johansen2007} which is an instability from a family of resonant drag instabilities \citep{Squire2018}. Its basic mechanism works as follows: radially drifting dust drags gas radially along, but angular momentum conservation causes the gas to increase its azimuthal velocity. This increased orbital gas speed drags dust forward in azimuth, accelerating it in the azimuthal direction and thus pushing it outward. This enhances the original dust enhancement and becomes a run-away effect. A detailed review of the SI related literature is beyond the scope of this review (see \citealp{Simon2022} for a recent review) and only the key findings are summarized in the following: While this linear phase of the classical streaming instability is reasonably well understood \citep{Squire2020}, the open questions lie in its interplay with other effects (external turbulence, particle size distributions) and in the clumping in the non-linear phase of the instability. 

It was found that the SI linear growth rate \citep{Auffinger2018,Umurhan2020} and strong clumping \citep{Gole2020} can be significantly reduced by viscosity. In contrast, \citet{Yang2018} showed that strong clumping can occur despite significant stirring in an MRI dead zone. This was confirmed by \citet{Xu2022} where moderate MRI turbulence levels even helped trigger the SI. Other mechanisms to drive turbulence, such as the VSI were shown to be able to coexist with the SI \citep{Schafer2020}. 

When particle size distributions are included, \citet{Krapp2019} found that the SI can be suppressed (see also  \citealp{Bai2010}, \citealp{Zhu2021} and \citealp{Paardekooper2020}). In contrast, \citet{Schaffer2021} find  that dust size distributions, although changing the dynamics of the system, do not suppress strong clumping, which the authors contribute to the fact that vertical settling leads in all cases to mid-plane concentrations where also linear analysis finds sufficiently high growth rates.

Our most current understanding (despite the caveats mentioned before, which clearly show that more work is required in the future), still remains that \emph{high} dust-to-gas ratios and \emph{large} particles are required to trigger strong clumping. What exactly \emph{high} and \emph{large} means was consolidated in \citet{Simon2022}, based on previous works by \citet{Carrera2015}, \citet{Yang2017}, and \citet{Li2021}, which is shown in the right panel of \autoref{fig:hill}. Dust to gas surface density ratios of  $\Sigd/\Sigg$ above 3\% and/or Stokes numbers above 0.01 are generally susceptible to strong clumping and dust evolution models allow the production of such sizes in pressure traps. In addition to that, the conditions for SI and strong clumping need to be retained long enough: \citet{Carrera2022}, for example, showed that a wide pressure bump in their simulation was not able to trigger strong clumping despite far exceeding the required criteria. As the authors state, this is likely due to the fact that the particles drift through the region of high concentration on time scales shorter than the growth timescale of the instability.

There do exist alternative proposals to concentrate particles to the onset of gravitational collapse: on small scales turbulent eddies can centrifugally eject particles, causing concentrations \emph{between} eddies, opposite to vortex trapping discussed in \autoref{sec:drift_mixing_equi}. This idea, proposed in \citet{Cuzzi2008}, was criticized as unlikely in more realistic models of turbulence by \citet{Pan2011}, but it remains the subject of ongoing discussions \citep{Hartlep2020,Klahr2020}. A second alternative is called the secular gravitational instability, where the self-gravity of a dust ring is aided by gas drag which can damp the dust eccentricy and enable continued growth of the initial perturbation \citep[e.g.][]{Ward2000,Takahashi2014}. Recent works on this mechanism includes \citet{Tominaga2020} and \citet{Abod2019} which show that the boundaries between the mechanisms get blurred in a pressure bump and that on small scales, the collapse is likely governed by the diffusing effects of turbulence \citep[see][]{Klahr2020,Gerbig2020}.

In summary, most models agree that accumulations of partially decoupled particles can trigger planetesimal formation. This means that some amount of particle growth is required, in order to provide particles large enough to participate in these mechanisms. Global models of dust evolution routinely produce particles significantly larger than the initially well coupled monomers. It is therefore likely that coagulation models (that mostly fail to produce planetesimals) and models of gravitational collapse of pebble-overdensities (that would not work with small dust grains) go hand-in-hand. In order to model planetesimal formation starting from small dust, therefore requires recipes that predict when the conditions for planetesimal formation (that is strong clumping) are reached. In doing so, a wide variety of models exist, see the discussion in \citet{Drazkowska2023}. The right panel in \autoref{fig:hill} presents the combined criteria of \citet{Li2021} and \citet{Simon2022} in comparison to a widely-used criterion which requires the mid-plane dust-to-gas ratio to reach unity \citep[][and others]{Drazkowska2016,Schoonenberg2017}. At the bottom, the ranges of Stokes numbers are displayed that are reached in fragmentation-limited growth for fragmentation velocities between 1 and $\SI{10}{m.s^{-1}}$. It can be seen that for a high metallicity, there always exists a range of Stokes numbers that can contribute to planetesimal formation. Such high metallicities $Z$ are however only reached through some form of dust-concentrating mechanism. This means that the planetesimal formation process can bee seen as a three stage-process:
\begin{itemize}
    \item Dust needs to grow via collisions to sizes that are partially decoupled from the gas, often called \textit{pebbles}. There is no fixed size associated with this process, it is rather understood as particles with a Stokes number $\St>\alpha$.
    \item Pebbles are susceptible to radial, vertical, and azimuthal drift and can therefore be efficiently trapped, or through radial drift efficiently be supplied to the inner disk regions. Some mechanism accumulates the dust beyond the percent-level. A range of potential mechanisms exist such as radial drift \citep[e.g.][]{Youdin2002}, pressure traps \citep[e.g.][and many others]{Whipple1972,Klahr1997,Rice2006,Pinilla2012} or dead-zone boundaries \citep[e.g.][]{Kretke2007,Lyra2008,Brauer2008b,Dzyurkevich2010,Pinilla2016b,Charnoz2021}, zonal flows \citep{Johansen2009,Bai2014}, the secular gravitational instability \citep[e.g.][]{Takahashi2014,Tominaga2020}, or traffic-jam effects at the water snow line \citep{Birnstiel2010,Saito2011} or outside of condensation fronts through sintering \citep{Okuzumi2016} or aided by dust back-reactions \citep{Drazkowska2017,Schoonenberg2017,Garate2020}.
    \item If the dust-to-gas ratio near the mid-plane approaches unity, and conditions for \emph{strong clumping} are fulfilled, pockets of high dust density can exceed the Hill density and further collapse to planetesimals is possible. This closes the gap between the high dust-to-gas ratio in the pressure trap (see \autoref{fig:hill}) and the even higher Hill density that is required for gravitational collapse.
\end{itemize}

At this point is worth revisiting the observations of disk sub-structure in disks: the optical depth at millimeter wavelength appears universally within the order of $\sim 0.4$, \citep{Cazzoletti2018,Isella2018,Dullemond2018b,Macias2019}. Given typical opacities (see \autoref{sec:obs_tracers}), and particle sizes, the fluxes suggest high local dust densities.  As shown in \citet{Stammler2019}, this universal optical depth within dust sub-structure is naturally explained if planetesimal formation truncates the dust-to-gas ratio beyond the critical value around unity. The effects of high albedo would only further increase the available dust mass for a given optical depth. Together with the fact that pressure perturbations are seen in gas kinematics \citep{Teague2018,Rosotti2020}, this is strongly suggesting that the observed rings are viable places for planetesimal formation. If true, such dust rings can also aid further planet formation by pebble accretion, as shown in \cite{Lau2022} and \cite{Jiang2023} who show that planetary cores can form rapidly in pebble rings (see also \citealp{Guilera2020}). Taken together, pressure maxima have the potential to trap dust particles (as already nicely depicted in \citealp{Whipple1972}), and provide the conditions for the dust to grow to sizes and Stokes numbers high enough for strong clumping to form planetesimals, while also acting as migration trap \citep{Morbidelli2020,Lau2022} for cores and at the same time as storage of dust for rapid pebble accretion. Thus, pressure maxima potentially solve many problems of planetesimal and planet formation all at once. Future studies will have to test whether this neat solution indeed works as currently envisioned and how substructure during the earliest stages of disk evolution form.

% -----------------------------------------------------------

% Summary Points
\begin{summary}[SUMMARY POINTS]
\begin{enumerate}
\item Observations over the last decade have transformed the field of protoplanetary disks and planet formation by enabling high-resolution imaging, large survey sizes, and gas kinematic studies of planet-forming disks. At millimeter wavelengths they revealed ubiquitous, mostly axi-symmetric sub-structures, in some cases showing asymmetries and spirals.
\item Observations suggest particle sizes of at least a few \SI{100}{\micro m} and significant dust accumulation in the rings. These features have been predicted by and can be explained with models of dust growth and transport.
\item Kinematics and direct detections indicate planets as the origin of many observed rings. This would imply that planet formation happens very early, well within a million years. Growth and trapping of dust in pressure traps can create the right conditions for rapid planetesimal formation and pebble accretion if turbulence levels are low enough and bouncing/fragmentation threshold velocities high enough. The observed, universally high optical depths of dust traps is strongly suggestive of a limiting effect such as the streaming instability setting in at high dust-to-gas ratios. With pressure traps being an attractive environment for planet formation and planets being the  most straight forward way to form pressure traps, one questions remains: Which one formed first, planet or pressure trap?
\item Strong asymmetries can mostly be explained by vortices and in some cases by binary companions. The wide range of structures observed in disks, as well as the strong variation in key parameters underline a wide diversity in disk properties and conditions. Particularly interesting are indications of widely varying turbulence levels, which is the most influential parameter of dust evolution models alongside the critical velocities for bouncing or fragmentation.
\end{enumerate}
\end{summary}

% Future Issues
\begin{issues}[FUTURE ISSUES]
    \begin{enumerate}
    \item Through dust optical properties and their wavelength-dependence, constraints on the maximum particle sizes are possible, but do not yet show a consistent picture. Tighter constraints on the composition and opacity of dust and the particle size distributions will be crucial for constraining the physics and conditions of early planet formation.
    \item Particle porosity remains a major unknown from both a theory and an observational standpoint. A more consistent picture of porosity evolution will be a key to understand dust evolution in planet forming disks.
    \item The apparently rapid formation of planets require more focus on the very early stages of disk formation and evolution, including infall (possibly asymmetrical streamers or late/sustained infall), early formation mechanisms of sub-structure and the effects of outbursts on the gas disk, the dust and their composition.
    \item Larger uniform samples that cover also small disks and disks with very low fluxes will be able to put tight constraints on disk population models, in particular the dissipation stages of disks.
    \item The origin of pressure traps and the dynamics within them remain subject to ongoing research. Observational methods to distinguish theoretical models are urgently required as observations probe deeper into the optically thick regions of disks.
    \item As already demonstrated by \citet{Varga2021}, \citet{Perraut2021}, or \citet{Perotti2023}, the inner regions of disks can be probed through interferometry or spectroscopy. In the future, this will enable to glimpse into the terrestrial planet forming regions \citep[see][]{Ricci2018} and probe dust evolution inward and around the snow line and provide crucial constraints on the disk dynamics and the formation mechanisms of terrestrial planets.
    \end{enumerate}
\end{issues}

% =========================================================
%Disclosure
\section*{DISCLOSURE STATEMENT}
The author is not aware of any affiliations, memberships, funding, or financial holdings that might be perceived as affecting the objectivity of this review. 

% Acknowledgements
\section*{ACKNOWLEDGMENTS}
I am grateful to many colleagues for helpful discussions and providing comments including Akimasa Kataoka, Sebastian Stammler, Thomas Henning, Hubert Klahr, Fabian Binkert, and in particular Chris Ormel for his detailed comments and furthermore to Rich Teague and Sean Andrews for sharing data.
This work made use of
\texttt{numpy} \citep{vanderWalt2011},
\texttt{matplotlib} \citep{Hunter2007},
\texttt{dustpy} \citep{Stammler2022},
\texttt{dsharp\_opac} \citep{Birnstiel2018},
\texttt{optool} \citep{optool},
\texttt{astropy} \citep{astropy},
\texttt{scipy} \citep{scipy}, and
\texttt{pandas} \citep{pandas}.
I acknowledge funding from the European Research Council (ERC) under the
European Union's Horizon 2020 research and innovation programme under grant
agreement No 714769 and funding by the Deutsche Forschungsgemeinschaft (DFG,
German Research Foundation) under grants
361140270,
325594231,
and Germany's Excellence Strategy - EXC-2094 - 390783311.

% References
%
% Margin notes within bibliography

\bibliographystyle{ar-style2}
\bibliography{bibliography}

\end{document}